\newif\ifpreprint
  \preprintfalse        % 12pt, single column, double spacing
  \preprinttrue         % for preprint form (double column, 10pt)
\newif\ifpdf
\ifpreprint
	\newif\ifastroph
	  \astrophfalse % full resolution figures
	  \astrophtrue  % bitmap at 72dpi (Fig 6 only)
        \documentclass[twoside,preprint2]{my_aastex}
        \ifx\pdfoutput\undefined\pdffalse\else\pdfoutput=1\pdftrue\fi
        \usepackage{amsmath,graphics,sects,times,multicol,hyperref}
\else
        \documentclass{aastex}
        \ifx\pdfoutput\undefined\pdffalse\else\pdfoutput=1\pdftrue\fi
        \usepackage{amsmath,graphics}
        \ifpdf
        \usepackage{times}
        \fi
\fi

\ifpreprint
\else
  \shorttitle{Modeling the disruption of the globular cluster Pal\,5 by
             Galactic tides}
  \shortauthors{Dehnen, Odenkirchen, Grebel, \& Rix}
\fi
\slugcomment{submitted for publication in the Astronomical Journal;
	     revised version}

\begin{document}

\ifpreprint \thispagestyle{empty} \fi
\title{MODELING THE DISRUPTION OF THE GLOBULAR CLUSTER PAL\,5 BY
	GALACTIC TIDES}
\author{Walter Dehnen\altaffilmark{123},
	Michael Odenkirchen\altaffilmark{3},
	Eva K.~Grebel\altaffilmark{43} and
	Hans-Walter Rix\altaffilmark{3}}
\ifpreprint\relax\else
%% package hyperref does interfere with these.
%% So we instead use \footnotetext after the first \section below.
\altaffiltext{1}{Department of Physics \& Astronomy,
              	 University of Leicester, Leicester LE1 7RH, United Kingdom}
\altaffiltext{2}{Astrophysikalisches Institut Potdsam, An der Sternwarte 16,
              	 D-14482 Potdsam, Germany}
\altaffiltext{3}{Max-Planck Institut f\"ur Astronomie, K\"onigstuhl 17,
              	 D-69117 Heidelberg, Germany}
\altaffiltext{4}{Astronomisches Institut der Universit\"at Basel,
	      	 Venusstrasse 7, CH-4102 Basel, Switzerland}
\fi
\email{walter.dehnen@astro.le.ac.uk;
       odenk@mpia.de;
       grebel@astro.unibas.ch;
       rix@mpia.de}

\begin{abstract} \ifpreprint\noindent\fi
  The globular cluster Pal\,5 is remarkable not only because of its
  extended massive tidal tails, but also for its very low mass and
  velocity dispersion and its size, which is much larger than its
  theoretical tidal radius. In order to understand these extreme
  properties, we performed more than 1000 $N$-body simulations of
  clusters traversing the Milky Way on the orbit of Pal\,5. Tidal shocks
  at disk crossings near perigalacticon dominate the evolution of
  extended low-concentration clusters, resulting in massive tidal tails
  and often in a quick destruction of the cluster. The overlarge size of
  Pal\,5 can be explained as the result of an expansion following the
  heating induced by the last strong disk shock $\sim150$\,Myr ago. Some
  of the models can reproduce the low observed velocity dispersion and
  the relative fractions of stars in the tails and between the inner and
  outer parts of the tails. Our simulations illustrate to which extent
  the observable tidal tails trace out the orbit of the parent
  object. The tidal tails of Pal\,5 show substantial structure not seen
  in our simulations. We argue that this structure is probably caused by
  Galactic substructure, such as giant molecular clouds, spiral arms,
  and dark-matter clumps, which was ignored in our modeling.

  Clusters initially larger than their theoretical tidal limit remain
  so, because, after being shocked, they settle into a new equilibrium
  near apogalacticon, where they are unaffected by the perigalactic
  tidal field. This implies that, contrary to previous wisdom, globular
  clusters on eccentric orbits may well remain super-tidally limited and
  hence vulnerable to strong disk shocks, which dominate their evolution
  until destruction. Our simulations unambiguously predict the
  destruction of Pal\,5 at its next disk crossing in
  $\sim$\,110\,Myr. This corresponds to only 1\% of the cluster
  lifetime, suggesting that many more similar systems could once have
  populated the inner parts of the Milky Way, but have been transformed
  into debris streams by the Galactic tidal field.

\end{abstract}

%%%%%%%%%%%%%%%%%%%%%%%%%%%%%%%%%%%%%%%%%%%%%%%%%%%%%%%%%%%%%%%%%%%%%%%%%%
\section{Introduction} \label{sec:intro}
%%%%%%%%%%%%%%%%%%%%%%%%%%%%%%%%%%%%%%%%%%%%%%%%%%%%%%%%%%%%%%%%%%%%%%%%%%
\ifpreprint
%% this is to replace \altaffiltext
\footnotetext[1]{Department of Physics \& Astronomy,
              	 University of Leicester, Leicester LE1 7RH, United Kingdom}
\footnotetext[2]{Astrophysikalisches Institut Potdsam, An der Sternwarte 16,
              	 D-14482 Potdsam, Germany}
\footnotetext[3]{Max-Planck Institut f\"ur Astronomie, K\"onigstuhl 17,
              	 D-69117 Heidelberg, Germany}
\footnotetext[4]{Astronomisches Institut der Universit\"at Basel,
	      	 Venusstrasse 7, CH-4102 Basel, Switzerland}
\setcounter{footnote}{4}
\fi
%%%%%%%%%%%%%%%%%%%%%%%%%%%%%%%%%%%%%%%%%%%%%%%%%%%%%%%%%%%%%%%%%%%%%%%%%%
Amongst the globular clusters of the Milky Way galaxy, Pal\,5 is
exceptional in many respects. First, it is one of the faintest and least
massive of these objects; its velocity dispersion is so small that it
could only recently be determined, using high-resolution spectroscopy,
to be below 1\,km\,s$^{-1}$ \citep[][ hereafter paper~I]{paper1}. Second
and most remarkable, it has been detected, using data from the Sloan
Digital Sky Survey \citep[SDSS; ][]{SDSS}, to possess a pair of tidal
tails, which extend at least 4\,kpc in opposite directions from the
cluster and contain more stars then the cluster itself \citep[][
hereafter paper~II]{paper0,Rocki,paper2}. Pal\,5 is actually the first
and only globular cluster so far for which such tails have been detected
at a comparable level of significance. Thus, this is the first direct
evidence that mass-loss induced by the Galactic tidal field can
substantially affect the evolution of globular clusters, a process which
is speculated to have destroyed many low-mass clusters of an initially
much richer Galactic system of clusters \citep[e.g.,][]{vh97}.

The tidal field generated by a smooth Galactic mass distribution
stretches the cluster in a direction towards the Galactic center and
weakly compresses it in tangential directions. The stretching creates
drains through which stars from the outer parts of the cluster are
carried away. This limits the bound part of a cluster which is in
equilibrium with the tidal field to a \emph{tidal radius},
$r_{\mathrm{tid}}$, whose size may be estimated by equating the
outwardly directed stretching tidal force to the cluster's internal
gravitational attraction. For a cluster of mass $M$ orbiting at
galactocentric radius $R$ in a galaxy with constant circular speed
$v_{\mathrm{c}}$, we get\footnote{In the literature this equation often
comes with an extra factor $1/2$, which occurs either if the orbit is
assumed to be circular or the Galactic potential to be that of a point
mass. If both assumptions are made, the factor is $1/3$.}
\begin{equation} \label{eq:rtid}
  r_{\mathrm{tid}}^3 \simeq \frac{GM}{v_{\mathrm{c}}^2}\,R^2.
\end{equation}
For eccentric orbits, $r_{\mathrm{tid}}$ varies along the orbit, and is
smallest at perigalacticon. The strong tidal force at perigalacticon
acts only for a short fraction of the orbital period, and passages of
the perigalacticon are often referred to as tidal shocks or `bulge
shocks'.

The tidal field generated by the stellar disk with its steep vertical
gradient is different. It is much stronger than that of the smooth halo
and bulge, and it is compressive. The latter is because, whilst the
cluster is crossing the disk, its stars feel the additional attraction
of the disk stars within the cluster. However, the strong compression
only acts for a short time. This allows to use the impulse
approximation, which gives for the velocity change of a star at vertical
distance $z$ from the cluster center
\citep{OSC72,Sp87}
\begin{equation} \nonumber
  \Delta v = -2g_{\mathrm{m}}\,z/V,
\end{equation}
where $V$ is the vertical velocity of the cluster crossing the disk and
$g_{\mathrm{m}}\approx2\pi\,G\,\Sigma(R)$ the maximal vertical
acceleration exerted by a (thin) disk with surface density
$\Sigma(R)$. These changes in velocity result in relative changes of the
specific energies of the stars of the order of
\begin{equation} \label{eq:dE}
  \frac{\langle|\Delta E|\rangle}{\langle|E|\rangle} \approx
  \frac{\sigma\,g_{\mathrm{m}}r_{\mathrm{h}}}{V\langle|E|\rangle}
  \approx \frac{2 g_{\mathrm{m}}r_{\mathrm{h}}}{\sigma\,V},
\end{equation}
where we have used $\langle|z|\rangle\approx r_{\mathrm{h}}/2$ with
$r_{\mathrm{h}}$ denoting the cluster half-mass
radius. Low-concentration systems tend to have large half-mass radii and
low velocity dispersions, which makes them very susceptible to disk
shocks, in particular at small galactocentric radii where $\Sigma(R)$
and hence $g_{\mathrm{m}}$ are large. Not surprisingly therefore,
low-concentration clusters are hardly found in the inner Galaxy, where
not only the shocks are much stronger but also occur much more
frequently than for orbits in the Galactic halo.

The velocity changes also result in an average heating per star of
$\langle\Delta E\rangle \approx 2g_{\mathrm{m}}^2r_{\mathrm{h}}^2/3V^2$
\citep{OSC72,Sp87}. One may use this result to estimate the ``shock
heating time'' $t_{\mathrm{sh}}$, the time scale over which disk shocks
significantly affect the cluster dynamics:
\begin{equation} \label{eq:tsh}
  t_{\mathrm{sh}} = t_{\mathrm{disk}}
	\frac{-\langle E\rangle}{\langle\Delta E\rangle}
	= t_{\mathrm{disk}} \frac{3}{4} \frac{\sigma^2V^2}
		{g_{\mathrm{m}}^2r_{\mathrm{h}}^2}
\end{equation}
\citep{glo99}. Here, $t_{\mathrm{disk}}$ is the time between strong disk
shocks, which for eccentric orbits, such as that of Pal\,5, equals the
orbital period $P$.

Apart from tidal forcing, the evolution of globular clusters is also
driven by internal processes, such as stellar mass loss, two-body
relaxation and binary interactions. These internal processes cause the
cluster to (eventually) undergo a core collapse and result in
mass-segregation and evaporation of preferentially low-mass stars. While
the evolution of isolated globular clusters has been the subject of many
studies, the combined effects of these internal and the external
processes have been rarely investigated. \cite{go97} used Fokker-Planck
simulations including two-body relaxation, tidal limitation and disk
and bulge shocks to investigate the dissolution of the individual
Galactic globular clusters (though their adopted mass for Pal\,5 is five
times larger than our best current value). With the same method,
\cite{glo99} studied the evolution of globular clusters with various
concentrations $c$, defined as
$c\equiv\log(r_{\mathrm{lim}}/r_{\mathrm{core}})$ from the cluster's
limiting and core radius, and ratios $\beta\equiv
t_{\mathrm{rh}}/t_{\mathrm{sh}}$ between the two-body relaxation time
$t_{\mathrm{rh}}$ and the shock heating time (\ref{eq:tsh}). They
considered clusters with $c\in[0.6,\,2.6]$ and $\beta \in
[10^{-5},\,10^2]$ orbiting the Milky Way on the orbit of the cluster
NGC\,6254 and concluded that (i) for $\beta\ga0.1$ disk shocks
dominate the cluster evolution and may lead to quick destruction, (ii)
for smaller values of $\beta$ tidal forcing accelerates the
two-body-relaxation driven evolution, but also that (iii), in response
to disk shocking, the cluster compacts, substantially reducing
$\beta$ and diminishing the further importance of disk shocks. Pal\,5
has $c\sim0.6$ (paper~II) and $\beta\sim10$
(section~\ref{sec:shocks}), which means that, according to Gnedin et
al., the evolution of Pal\,5 is entirely dominated by disk shocks.

Collisional $N$-body-simulation studies which investigate internal
processes along with a time varying tidal field either ignore disk
shocks \citep{bm03} or account for disk shocks but assume an otherwise
constant tidal field \citep{vh97}.

Unfortunately, neither of these studies is directly applicable to
Pal\,5, since they either exclude relevant external processes or do not
cover the globular-cluster parameters relevant for Pal\,5. Gnedin et
al., for example, assumed in their study that disk shocks are slow and
rare, i.e.\ the cluster-internal dynamical time is shorter than the
duration of the shock and much shorter than $t_{\mathrm{disk}}$, whereas
for Pal\,5 shocks are fast and frequent (see
section~\ref{sec:shocks}). Also, the orbit of Pal\,5 differs
significantly from that of NGC\,6254, as adopted by these authors.

Moreover as far as we are aware, all studies of globular cluster
evolution start with clusters limited by their (perigalactic) tidal
radius, assuming that this limitation is quickly achieved by the tidal
force field. As we will see below, this assumption is not justified,
neither theoretically as our simulations will demonstrate, nor
observationally, since Pal\,5 is an excellent counterexample, see
section~\ref{sec:extent}.

The primary goal of the present paper is to study, via detailed
$N$-body simulations, the effect of Galactic tides on a globular cluster
moving on the orbit of Pal\,5 and to quantitatively compare the
observable properties of the models with those of the cluster in an
attempt to understand its current dynamical state and to constrain the
formation history of this object. As a byproduct, our simulations will
also give valuable insights into the evolution of low-concentration
globular clusters with $\beta\gg1$ experiencing strong disk shocks in
conjunction with a time varying tidal field.

In section~\ref{sec:prop} we summarize the observed structural and
dynamical properties of Pal\,5. The $N$-body simulations are presented
in section~\ref{sec:simul}. The dynamics of the simulated tidal tails
are discussed in detail in section~\ref{sec:tail}. In
section~\ref{sec:compare}, we compare our simulations directly to the
data for Pal\,5. The results of the $N$-body simulations and the
implication for Pal\,5 are discussed, respectively, in
sections~\ref{sec:disc:I} and \ref{sec:disc:II} and summarized in
section~\ref{sec:summ}.

%%%%%%%%%%%%%%%%%%%%%%%%%%%%%%%%%%%%%%%%%%%%%%%%%%%%%%%%%%%%%%%%%%%%%%%%%%
\section{The Observed Properties of Pal\,5} \label{sec:prop}
%%%%%%%%%%%%%%%%%%%%%%%%%%%%%%%%%%%%%%%%%%%%%%%%%%%%%%%%%%%%%%%%%%%%%%%%%%
We shall now summarize the structural and kinematic properties of the
globular cluster Pal\,5, which are relevant for the present study. Most
of these properties have been presented and are discussed in
papers~I\,\&\,II.

%%%%%%%%%%%%%%%%%%%%%%%%%%%%%%%%%%%%%%%%
\ifpreprint
  \begin{figure}[t]
    \centerline{\resizebox{60mm}{!}{\includegraphics{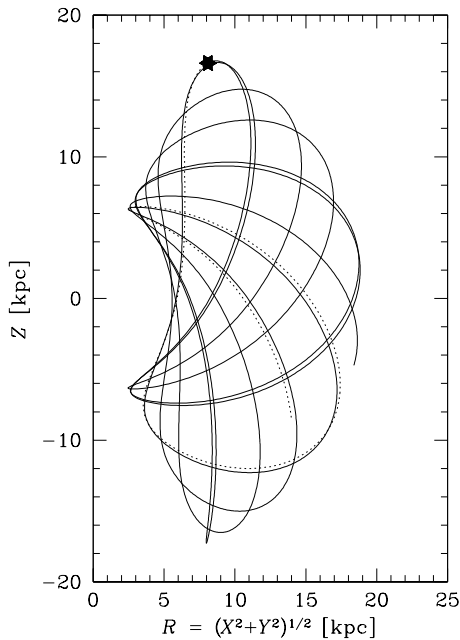}}}
	\figcaption{Orbit of Pal\,5, projected into the meridional
	plane, for the assumed Galactic potential. The current position
	of Pal\,5 is indicated by the star; the trajectory is plotted
	for the last 2.95\,Gyr (\emph{solid}), corresponding to 10
	radial periods, and the future 0.5\,Myr (\emph{dotted}).
	\label{fig:orbit}}
  \end{figure}
\else
  \placefigure{fig:orbit}
\fi
%%%%%%%%%%%%%%%%%%%%%%%%%%%%%%%%%%%%%%%%%%%%%%%%%%%%%%%%%%%%%%%%%%%%%%%%%%
\subsection{The Orbit} \label{sec:orbit}
As shown in paper~II, the orbit of Pal\,5 is rather tightly constrained
by its position and radial velocity in conjunction with the observed
orientation and curvature of the tidal tail. This is the case, because
the tail stars deviate from the cluster orbit only very little, as is
evident, for instance, for the simulated tidal tail in
Figure~\ref{fig:tail:morph} below.  Figure~\ref{fig:orbit} shows the
meridional projection of that orbit obtained with our standard model for
the Galactic gravitational potential (see section~\ref{sec:simul:pot}
for details).  Different assumptions for the Galactic potential, e.g.\
different circular velocities, result in very similar orbits, except for
their periods (paper~II).  In particular, the perigalactic and
apogalactic radii are always at about $5.5$ and $19\,$kpc.

%%%%%%%%%%%%%%%%%%%%%%%%%%%%%%%%%%%%%%%%%%%%%%%%%%%%%%%%%%%%%%%%%%%%%%%%%%
\subsection{Mass and Velocity Dispersion} \label{sec:mass_sigma}
With a total luminosity of only $M_V=-4.77\pm0.20$ (paper~I) Pal\,5 is
one of the faintest Galactic globular clusters. It also has an
unusually flat stellar luminosity function \citep{GS01} implying that
its mass-to-light ratio is atypically low. Together, this let us
estimate in paper~I the total mass to be $5200\pm700\,M_\odot$,
substantially less than previous estimates.

In paper~I, we measured the line-of-sight velocity dispersion
$\sigma_{\mathrm{los}}$ from high-resolution spectra of 18 giant stars
within 6\arcmin\ of the cluster center to be at most 1.1\, km\,
s$^{-1}$. However, the line-of-sight velocity distribution is
significantly non-normal but has extended high-velocity tails, which
dominate the calculation of $\sigma_{\mathrm{los}}$. When modeling this
distribution as Gaussian for the cluster dynamics plus a contribution
from binaries to account for the tails, we obtained for the dynamically
relevant velocity dispersion $\sigma_{\mathrm{los}} =
0.22^{+0.19}_{-0.10}\,$km\,s$^{-1}$.

As we discussed in paper~I, these estimates for the mass and velocity
dispersion are consistent with the best-fit King model and the
assumption of dynamical equilibrium. However, this does not imply that
the cluster is in equilibrium and we shall now see that it cannot
possibly be.

%%%%%%%%%%%%%%%%%%%%%%%%%%%%%%%%%%%%%%%%
\ifpreprint
  \begin{figure}[t]
    \centerline{\resizebox{80mm}{!}{\includegraphics{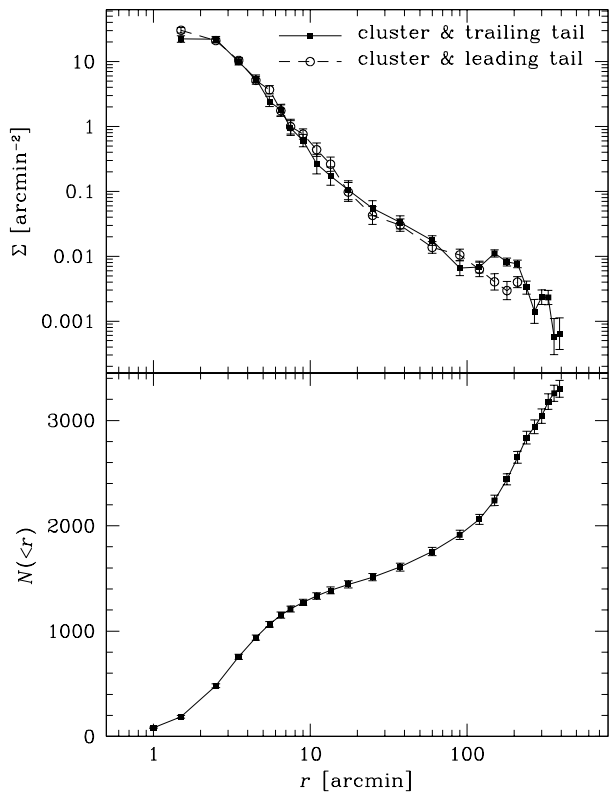}}}
    \figcaption{Radial profile of the azimuthally averaged surface
    number density of SDSS stars in Pal\,5 and its tails as measured in
    paper~II (\emph{top}), and cumulative number distribution averaged
    over both tails (\emph{bottom}). The typical mass of Pal\,5 stars in
    SDSS is about 0.8\,$M_\odot$.\label{fig:surf}}
  \end{figure}
\else
  \placefigure{fig:surf}
\fi
%%%%%%%%%%%%%%%%%%%%%%%%%%%%%%%%%%%%%%%%%%%%%%%%%%%%%%%%%%%%%%%%%%%%%%%%%%
\subsection{Extent and Tidal Radius} \label{sec:extent}
Figure~\ref{fig:surf} shows the surface density, averaged over annuli,
of the cluster and the tidal tail. The profile of the cluster itself
(which can easily be obtained from data in directions perpendicular to
the tail) appears to be truncated at a limiting radius of about
$16^\prime\!.1\approx107\,$pc, see Fig.~5 of paper~II. This radius must
be compared to the theoretical tidal radius (equation \ref{eq:rtid}) for
the cluster, which for $v_{\mathrm{c}}=220\,$km\,s$^{-1}$ and the mass
estimate given above is $\sim54\,$pc at the current location of Pal\,5
and only $\sim24\,$pc at the perigalacticon of its orbit. This latter
value is much smaller than $r_{\mathrm{lim}}$ and actually equals the
cluster's present-day core radius.

Thus, the cluster is much larger than its tidal radius and cannot
possibly be in equilibrium with the Galactic tidal force
field\footnote{In previous studies, this discrepancy was not detected
because (i) the mass of the cluster was overestimated by a factor of 4-6
and (ii) the perigalactic radius was unknown.}. This is surprising and
shows that the current dynamical state of Pal\,5 must be a peculiar
one. In particular, any estimation which is based on the assumption of
equilibrium may well be in error.

%%%%%%%%%%%%%%%%%%%%%%%%%%%%%%%%%%%%%%%%%%%%%%%%%%%%%%%%%%%%%%%%%%%%%%%%%%
\subsection{The Importance of Disk Shocks} \label{sec:shocks}
After having assessed the various properties of Pal\,5 and its orbit, we
may now consider the importance of disk shocks at disk crossings. The
tidal force acting on a star at position $\boldsymbol{r}$ with respect
to the cluster center is given by
\begin{equation} \label{eq:Ftid}
  \boldsymbol{F}_{\mathrm{tid}}(\boldsymbol{r},t) =
  (\boldsymbol{r}\cdot\boldsymbol{\nabla})\,
   \boldsymbol{\nabla}\Phi(\boldsymbol{R}(t),t)
\end{equation}
where $\Phi$ is the gravitational potential of the Milky Way and
$\boldsymbol{R}(t)$ the galactocentric position of the cluster center
at time $t$. Equating $|\boldsymbol{F}_{\mathrm{tid}}|$ to the cluster
internal attraction $GM/r^2$ gives relation (\ref{eq:rtid}) for the
radius $r$ if $\Phi$ is taken to be the potential of a singular
isothermal sphere.

%%%%%%%%%%%%%%%%%%%%%%%%%%%%%%%%%%%%%%%%
\ifpreprint\relax \else
  \placefigure{fig:tfield}
\fi
%%%%%%%%%%%%%%%%%%%%%%%%%%%%%%%%%%%%%%%%
In Figure~\ref{fig:tfield}, we plot the strength of the tidal force
field, quantified by the eigenvalues of $\partial^2\Phi/\partial
x_i\partial x_j$ as function of time on Pal\,5's orbit in our model for
the Galactic potential (see section \ref{sec:simul:pot}). On top of a
smooth underlying tidal field, whose dominant effect is a stretching
(the absolute largest eigenvalue is negative, i.e.\ \emph{dotted\/} in
Fig.~\ref{fig:tfield}) and whose strength varies by a factor of $\sim10$
between apogalacticon and perigalacticon, there are short spikes
coinciding with disk crossings, when the tidal field is compressive and
$\ga10$ times stronger than otherwise. The duration of these shocks is
given by the disk scale height divided by the cluster's vertical
velocity and amounts to 10\,Myr or less. The strongest shocks, which
occur at disk crossings near perigalacticon, happen roughly once per
orbit.  Actually, the last such shock occurred $146\,$Myr ago, while the
next one is due in $110\,$Myr.

%%%%%%%%%%%%%%%%%%%%%%%%%%%%%%%%%%%%%%%%
\ifpreprint
  \begin{figure}[t]
  \centerline{\resizebox{80mm}{!}{\includegraphics{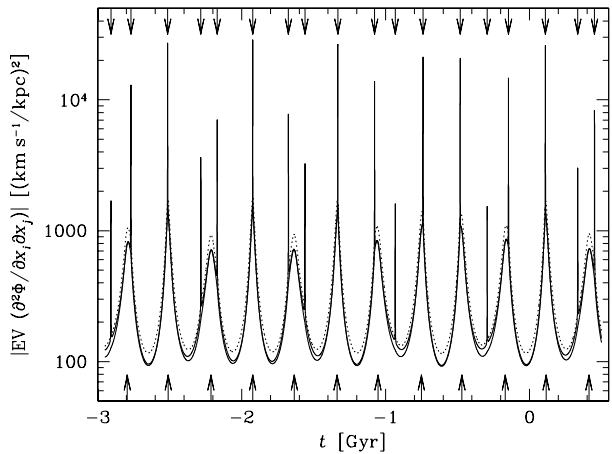}}}
  \figcaption{Strength of the tidal field along the orbit of Pal\,5
  (Fig.~\ref{fig:orbit}) as measured by the eigenvalues of
  $\partial^2\Phi/\partial x_i\partial x_j$. Positive (\emph{solid}) and
  negative (\emph{dotted}) eigenvalues correspond to compressive and
  stretching tidal forces, respectively. The little arrows on the top
  and bottom of the plot indicate moments of disk crossing and
  pericentric passages, respectively. $t=0$ corresponds to
  today.\label{fig:tfield}} \end{figure}
\fi
%%%%%%%%%%%%%%%%%%%%%%%%%%%%%%%%%%%%%%%%
With numbers collected in the previous subsections, we find from
equation~(\ref{eq:dE}) that a typical relative change of kinetic energy
amounts up to $\sim30\%$ for a disk shock near perigalacticon, while
equation (\ref{eq:tsh}) gives $t_{\mathrm{sh}}\sim2\,$Gyr for the
characteristic shock-evolution time scale.

Another quantity of interest is the internal dynamical time, which we
may estimate from the velocity dispersion to currently be
$\approx80\,$Myr at the half-mass radius $r_{\mathrm{h}}=30\,$pc. After
a strong disk shock, the cluster needs a few dynamical times to settle
into a new equilibrium. Since the dynamical time is not substantially
shorter than the time between subsequent shocks we anticipate that the
cluster hardly ever is in dynamical equilibrium.

Yet another important time scale is the two-body relaxation time
$t_{\mathrm{rh}}$, which we can estimate using Spitzer \& Hart's (1971)
formula
\citep[][ eq.~8.71]{BT87}.  At the half-mass radius, we find
\begin{equation}
  t_{\mathrm{rh}}
  \sim 20\,\mathrm{Gyr}\,
  \bigg[\frac{\sigma}{\scriptstyle 0.22\,\mathrm{km\,s^{-1}}}\bigg]^3 
  \left[\frac{\rho(r_{\mathrm{h}})}
    {\scriptstyle 0.01M_\odot\mathrm{pc^{-3}}}\right]^{-1}
  \left[\frac{M_\ast}{\scriptstyle 0.3M_\odot}\right]^{-1},
\end{equation}
which is too long for relaxation related processes to currently play an
important role and implies $\beta\sim10$ for Gnedin et al.'s
shock-importance parameter.

%%%%%%%%%%%%%%%%%%%%%%%%%%%%%%%%%%%%%%%%
\ifpreprint

\begin{table}[t]
  \footnotesize
  \refstepcounter{table} \label{tab:times}
  \begin{center}
    {\scshape
      Table \arabic{table} \\
      Time Scales for Pal\,5
    } \\[1ex]
  \begin{tabular}{llr@{$\,$}r@{$\,$}l}
    \hline\hline
    \\[-1.5ex]
    \multicolumn{1}{c}{dynamical process} &
    \multicolumn{4}{c}{time scale}
    \\[0.3ex]
    \hline
    \\[-2ex]
    duration of shocks     &                   &$\la$    & 10&Myr\\
    internal dynamics      &$t_{\mathrm{dyn}}$ &$\sim$   & 80&Myr\\
    time between shocks    &$t_{\mathrm{disk}}$&$\approx$&300&Myr\\
    shock-driven evolution &$t_{\mathrm{sh}}$  &$\sim$   &  2&Gyr\\
    two-body relaxation    &$t_{\mathrm{rh}}$  &$\sim$   & 20&Gyr\\[0.3ex]
    \hline
  \end{tabular}\par \vspace{1ex}
  \begin{minipage}{58mm}{\hspace{1em} {\scshape Note.---}
    Cluster internal time scales are estimated at the half-mass radius
    $r_{\mathrm{h}}\approx30\,$pc.}
  \end{minipage}
  \end{center}
\end{table}

\else
  \placetable{tab:times}
\fi
%%%%%%%%%%%%%%%%%%%%%%%%%%%%%%%%%%%%%%%%
We have summarized these various time scales in
Table~\ref{tab:times}. Note that the estimates for internal time scales
($t_{\mathrm{dyn}}$, $t_{\mathrm{sh}}$, $t_{\mathrm{rh}}$) must be
regarded as crude ones, because they are based on the assumption of
dynamical equilibrium, which we just saw is not really satisfied.  We
conclude, nonetheless, that disk shocks for Pal\,5 are fast (shock
duration $<t_{\mathrm{dyn}}$) and frequent ($t_{\mathrm{dyn}}\not\ll
t_{\mathrm{disk}}$) and entirely dominate the evolution of this object
($t_{\mathrm{sh}}<t_{\mathrm{rh}}$).
%%%%%%%%%%%%%%%%%%%%%%%%%%%%%%%%%%%%%%%%%%%%%%%%%%%%%%%%%%%%%%%%%%%%%%%%%%

%%%%%%%%%%%%%%%%%%%%%%%%%%%%%%%%%%%%%%%%%%%%%%%%%%%%%%%%%%%%%%%%%%%%%%%%%%
\section{The Simulations}\label{sec:simul}
%%%%%%%%%%%%%%%%%%%%%%%%%%%%%%%%%%%%%%%%%%%%%%%%%%%%%%%%%%%%%%%%%%%%%%%%%%
In order to model the tidal disruption of low-concentration globular
clusters in general and Pal\,5 in particular, we performed collision-less
$N$-body simulations, i.e.\ with softened gravity, suppressing
short-range interactions. This is entirely justified, because the
evolution of these systems is dominated by disk shocks and effects
driven by two-body relaxation (partly caused by close encounters) are
much less important (stellar mass loss affects clusters only in the
first Gyr of their life).

\subsection{The Galactic Gravitational Field}\label{sec:simul:pot}
%%%%%%%%%%%%%%%%%%%%%%%%%%%%%%%%%%%%%%%%%%%%%%%%%%%%%%%%%%%%%%%%%%%%%%%%%%
In our simulations, we considered only one choice for the Galactic
gravitational potential; any effects due to variations of the Galactic
potential are beyond the scope of this paper, but see the discussion in
section~\ref{sec:disc:pal5:tail}. Because the evolution of the simulated
cluster is dominated by disk shocks, it is important to model the
Galactic disk as realistically as possible. We used model 2 of
\cite{MM}, which contains a compound of three exponential disks for the
thick and thin stellar disk and for the ISM, respectively, and two
spheroidal components for the bulge and halo. The potential parameters
have been determined to fit all observational constraints known in 1998
and has a disk scale length of 2.4\,kpc, consistent with recent studies
of the infrared background light distribution in the Milky Way
\citep{DS01}.

\subsection{Initial Conditions}\label{sec:simul:init}
%%%%%%%%%%%%%%%%%%%%%%%%%%%%%%%%%%%%%%%%%%%%%%%%%%%%%%%%%%%%%%%%%%%%%%%%%%
To create suitable initial conditions, we used King models
\citep{king1,king2,king3} with barycenter at the position and velocity
of the orbit of Pal\,5 (see section~\ref{sec:orbit} and
Fig.~\ref{fig:orbit}) ten radial orbital periods (corresponding to
2.95\,Gyr) ago. King models have three free parameters, two scales (size
and mass) and a shape parameter $W_0$, which is a dimensionless measure
for the depth of the gravitational potential and equivalent to the
concentration $c$. As independent parameters, we used $W_0$, the total
initial mass $M_0$, and the galactocentric radius
\begin{equation}
  R_{\mathrm{t}} \equiv \sqrt{\frac{r_{\mathrm{lim}}^3}{G\,M_0}}
	\;v_{\mathrm{c}}
\end{equation}
with $v_{\mathrm{c}}=220\,$km\,s$^{-1}$, which means that the limiting
radius\footnote{The outermost radius of a King model is often called its
`tidal radius' with the idea that a globular cluster ought to be tidally
limited.  However, as Pal\,5 clearly demonstrates, the limiting radius
may be different, actually larger, than the theoretical tidal radius,
see section~\ref{sec:extent}.}  $r_{\mathrm{lim}}$ of the model equals
its tidal radius $r_{\mathrm{tid}}$ (equation~\ref{eq:rtid}) when at a
distance $R_{\mathrm{t}}$ from the Galactic center. With the choice of
$R_{\mathrm{t}}$ as parameter (rather than, say, $r_{\mathrm{lim}}$) the
importance of the tidal field is largely independent of the cluster mass
when the other two parameters are kept fixed.

We considered 1056 models from a grid of points in the 3D parameter
space. For $R_{\mathrm{t}}$, we took twelve values between 7.5 and
13\,kpc at steps of 0.5\,kpc. These are all larger than the perigalactic
radius of Pal\,5, guaranteeing that tidal interactions are
important. For $W_0$, we used eleven values between 1.75 and 4.25 in
steps of 0.25, corresponding to low concentrations $c$ between 0.46 and
0.88.  Finally, for the initial mass the eight values 8, 10, 12, 16, 20,
24, 28, and 32 thousand Solar masses have been employed.

%%%%%%%%%%%%%%%%%%%%%%%%%%%%%%%%%%%%%%%%
\ifpreprint
  \begin{figure*}[t]
    \centerline{\resizebox{175mm}{!}{\includegraphics{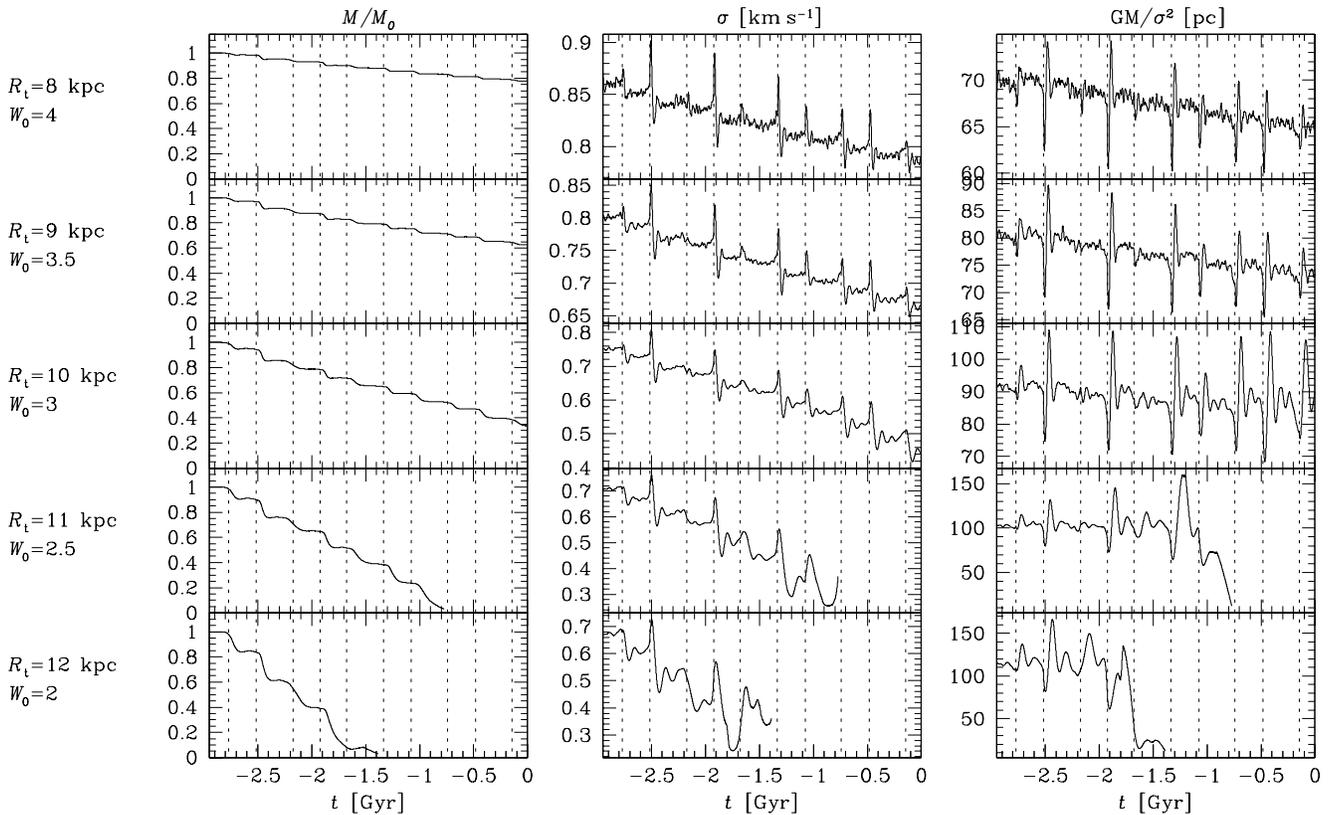}}}
    \figcaption{Time evolution of the mass (\emph{left}), velocity
    dispersion (\emph{middle}), and virial radius (\emph{right}) of
    the simulated globular cluster as function of time for five
    representative models with parameter $R_{\mathrm{t}}$ and $W_0$ as
    indicated and $M_0=12000\,M_\odot$. The \emph{dotted\/} vertical lines
    indicate disk crossings where the the tidal force is exceptionally
    large, quantified by the absolute largest eigenvalue of
    $\partial^2\Phi/\partial x_i\partial x_j$ exceeding
    7000\,(km\,s$^{-1}$\,kpc$^{-1}$)$^2$, see Fig.~\ref{fig:tfield}.
    The two simulations with $(R_{\mathrm{t}},W_0)=(11,2.5)$ and $(12,2)$
    have been stopped before $t=0$, because the number of bodies in the
    cluster dropped below 1000. \label{fig:evolv}}
  \end{figure*}
\fi
%%%%%%%%%%%%%%%%%%%%%%%%%%%%%%%%%%%%%%%%
Clearly, these initial conditions are, as often with $N$-body
simulations, somewhat artificial. We do not mean that 3\,Gyr ago Pal\,5
was well fit by a King model, rather we expect that it might already
have undergone severe tidal disturbances. However, the tidal field is
likely to quickly erase any specialities of our initial models, except,
of course, the lack of possible tidal tails originating from earlier
epochs. Moreover, even if one were willing to improve on the initial
conditions by integrating for the whole lifetime of a cluster, one had
to know the detailed history of the Galactic tidal field and would
require much more CPU time without gaining much scientific significance.

\subsection{Technical Details}\label{sec:simul:tech}
%%%%%%%%%%%%%%%%%%%%%%%%%%%%%%%%%%%%%%%%%%%%%%%%%%%%%%%%%%%%%%%%%%%%%%%%%%
We used the publicly available $N$-body code \textsf{gyrfalcON}, which
is based on Dehnen's (2000, 2002) force solver \textsf{falcON}, a tree
code with mutual cell-cell interactions and complexity
$\mathcal{O}(N)$. \textsf{falcON} not only conserves momentum exactly,
but also is about 10 times faster than an optimally coded Barnes \& Hut
(1986) tree code.

Each simulation used $N$=32000 bodies and a softening length of
$\epsilon=3\,$pc with what we call the P$_2$ softening kernel, i.e.\ the
Newtonian Greens function $\Phi=-G/r$ was replaced by
\[
  \Phi(r) = - \frac{G}{\sqrt{r^2+\epsilon^2}} \left[1+
    \frac{\epsilon^2}{2(r^2{+}\epsilon^2)} +
    \frac{3\epsilon^4}{4(r^2{+}\epsilon^2)^2}
  \right] \nonumber.
\]
The density of this kernel falls off more steeply at large $r$ than that
of the standard Plummer kernel and becomes actually negative (though
with absolute value smaller than for the Plummer kernel) such that force
bias is substantially reduced, see also Dehnen (2001).  With these
softening parameters, the maximum possible force between two bodies is
equal to that for Plummer softening with $\epsilon\approx1.4\,$pc, which
hence might be called the `equivalent Plummer softening length'.

The time integration was performed either for 2.95\,Gyr, i.e.\ until
today, or until cluster destruction. In practice, a cluster was
considered destroyed if the number of bodies within one initial limiting
radius from the cluster center dropped below $1000\simeq3\%$ of the
initial number. We used the leap-frog integrator with step size of
$2^{-11}\,\mathrm{Gyr}\approx0.5\,$Myr and a block-step scheme that
allowed up to 8 times smaller steps. The individual step sizes $\tau$
where adjusted in an almost time-symmetric way such that on average
$\tau=a^{-1}\,$kpc\,Gyr$^{-1}$, where $a$ denotes the modulus of the
acceleration. This ensured that disk crossings, which have a duration of
only a few Myr, are accurately integrated. With these settings, the
energy of an isolated cluster was conserved to 0.8\% over the period of
2.95\,Gyr. One full simulation corresponds to about 6000 block steps and
required about 75\,min of CPU time on a linux PC (AMD, 1800\,MHz). 1190
hours of CPU time were spent on the whole set of 1056 simulations, 361
of which were halted before $t=0$, because the cluster was found to be
dissolved.

%%%%%%%%%%%%%%%%%%%%%%%%%%%%%%%%%%%%%%%%
\ifpreprint\relax\else
  \placefigure{fig:evolv}
\fi
%%%%%%%%%%%%%%%%%%%%%%%%%%%%%%%%%%%%%%%%
\subsection{Cluster Evolution in the Tidal Field}\label{sec:simul:evolv}
%%%%%%%%%%%%%%%%%%%%%%%%%%%%%%%%%%%%%%%%%%%%%%%%%%%%%%%%%%%%%%%%%%%%%%%%%%
For five representative simulations, ranging from medium concentration
and small extent to low concentration and large extent, i.e.\ from least
to most vulnerable to Galactic tides, Figure~\ref{fig:evolv} shows the
time evolution of the cluster mass $M$, velocity dispersion $\sigma$,
and the ratio $GM/\sigma^2$.  They have been computed from those stars
that are within the original limiting radius from the cluster center,
which was determined iteratively as barycenter of the same stars
starting the iteration with the set of stars from the previous time
step.

\subsubsection{The Mechanics of Tidal Disk Shocks}
\label{sec:simul:evolv:mech}
%%%%%%%%%%%%%%%%%%%%%%%%%%%%%%%%%%%%%%%%%%%%%%%%%%%%%%%%%%%%%%%%%%%%%%%%%%
As is evident from this Figure, the cluster evolution is driven by disk
shocks (the instants of the strongest of which are indicated by
\emph{dotted\/} vertical lines), which may actually quickly destroy the
cluster, depending on its initial state.  Each strong disk shock causes
an almost instantaneous increase of the cluster's velocity dispersion
(middle panels of Fig.~\ref{fig:evolv}) by 8--25\%. This corresponds to
an increase of the cluster kinetic energy by 16--50\% and pushes the
cluster out of virial equilibrium. Stars which have been accelerated
beyond their escape velocity are lost from the cluster, resulting in a
drop of the cluster mass (left panels of Fig.~\ref{fig:evolv}), which is
delayed from the instant of the shock by the time needed for the stars
to drift out of the cluster.

In response to the heating by the disk shock, the cluster also
expands. This together with the loss of the fastest stars results in a
substantial drop of the velocity dispersion by approximately twice the
amount of the initial increase.  This drop occurs roughly on a dynamical
time scale, i.e.\ slower for the less concentrated and/or more extended
clusters. Subsequently, the velocity dispersion shows damped
oscillations until the cluster settles into a new equilibrium with
velocity dispersion lower than before the shock. The damping is due to
the fact that stars oscillate at different frequencies and is weakest
for low-concentration clusters, since for those the range in orbital
frequencies is smallest. For clusters of low concentration and/or large
extent, the settling into a new equilibrium is forestalled by the next
disk shock, i.e.\ these clusters are never in a state of dynamical
equilibrium.

%%%%%%%%%%%%%%%%%%%%%%%%%%%%%%%%%%%%%%%%
\ifpreprint
  \begin{figure}[t]
    \centerline{\resizebox{\columnwidth}{!}{\includegraphics{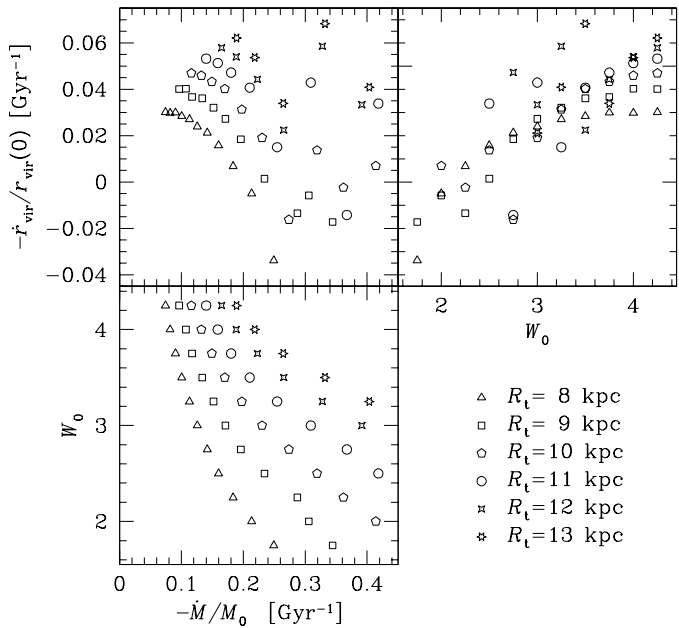}}}
    \figcaption{Relative mass-loss and shrinking rates plotted versus
    each other (\emph{top left}) and versus $W_0$ for model clusters
    that have not been dissolved after 2\,Gyr (for models dissolved
    earlier the uncertainties in the shrinking rate can be
    considerable). $\dot{M}$ and $\dot{r}_{\mathrm{vir}}$ have been
    obtained by straight-line fits over the entire time interval,
    excepting the 0.5\,Gyr before cluster dissolution, if
    applicable. For clarity, only models with $M_0=24000\,M_\odot$ are
    shown (other values for $M_0$ yield very similar results). The
    symbols refer to different values for $R_{\mathrm{t}}$ as indicated.
    \label{fig:rates}}
  \end{figure}
\else
  \placefigure{fig:rates}
\fi
\subsubsection{Mass Loss and Time Evolution}
%%%%%%%%%%%%%%%%%%%%%%%%%%%%%%%%%%%%%%%%%%%%%%%%%%%%%%%%%%%%%%%%%%%%%%%%%%
A remarkable observation from Fig.~\ref{fig:evolv} is the fact that the
orbit-averaged cluster mass decreases almost \emph{linearly\/} with time
over the durations simulated\footnote{We found the second time derivate
$\ddot{M}$ to be small in the sense that $|\dot{M}/\ddot{M}|>7\,$Gyr
with no preference for a sign.}. This implies that, at least for
low-concentration clusters, disk-shocking induced mass loss quickly
destroys the cluster. The orbit-averaged mass-loss rate depends strongly
on the size, but also on the concentration of the cluster, spanning one
order of magnitude in $\dot{M}/M_0$ amongst our simulations, see the
bottom left panel of Fig.~\ref{fig:rates}.

%%%%%%%%%%%%%%%%%%%%%%%%%%%%%%%%%%%%%%%%
\ifpreprint
  \begin{figure*}[t]
    \ifastroph
       \centerline{\resizebox{\textwidth}{!}{
	           \includegraphics{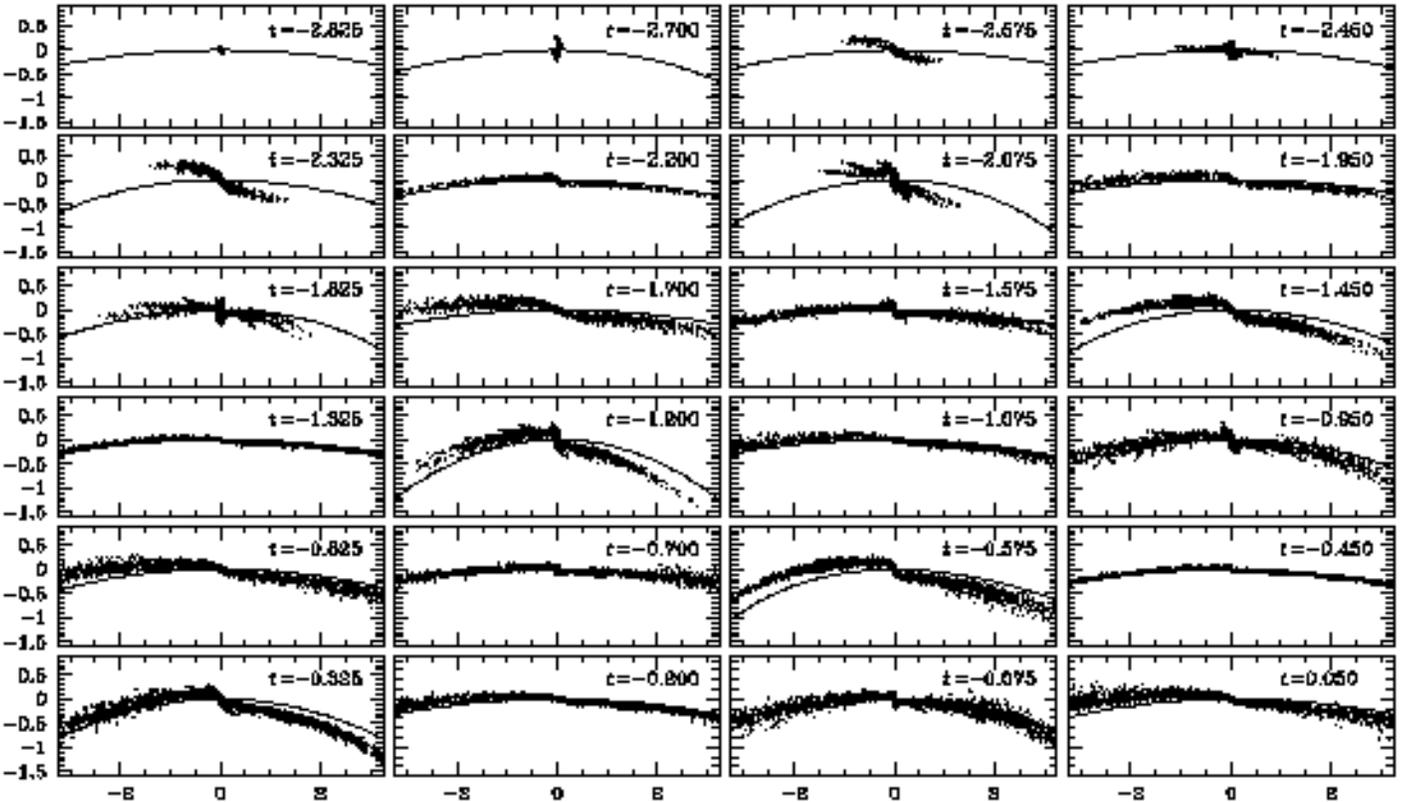}}}
    \else
       \centerline{\resizebox{\textwidth}{!}{\includegraphics{Dehnen_fig06}}}
    \fi
    \figcaption{Morphology of the simulated tidal tail in the simulation
    with parameters $R_{\mathrm{t}}=10\,$kpc, $W_0=2.75$, and
    $M_0=20000\,M_\odot$ (also called `model A') at various times, given
    in Gyr in each panel. For every tenth body, we plot the position (in
    kpc) projected onto the instantaneous orbital plane of the cluster,
    whose motion is towards the right in each panel.  The thin line
    indicates the cluster orbit.\label{fig:tail:morph}}
  \end{figure*}
\fi
%%%%%%%%%%%%%%%%%%%%%%%%%%%%%%%%%%%%%%%%
The linear mass loss is accompanied by an average decrease in velocity
dispersion. In the right panels of Fig.~\ref{fig:evolv}, we plot the
virial radius
\begin{equation}
  r_{\mathrm{vir}} \equiv GM/\sigma^2
\end{equation}
of the stars within the initial limiting radius of the cluster\footnote{
One might instead want to use only stars that are bound to the cluster
in the presence of the tidal field. However, due to strong changes of
the tidal field during shocks, a star may switch from bound to unbound
and back, resulting in spurious variations of the estimated
quantities. Moreover, the contamination by unbound stars is small and
does not significantly affect our estimates.}.  For most simulated
clusters that survive until the present time, we find a modest decrease
of the virial radius. In the top right panel of Figure~\ref{fig:rates},
we plot the relative shrinking rate, $-\dot{r}_{\mathrm{vir}}
/r_{\mathrm{vir}}(t=0)$, versus $W_0$ for various values of
$R_{\mathrm{t}}$ but fixed $M_0$ (the dependence on $M_0$ is very weak
at fixed $R_{\mathrm{t}}$ and $W_0$). Most model clusters shrink
slightly in response to the disk shocks with a rate that is primarily a
function of $W_0$, or, equivalently, of the initial cluster
concentration. Only very low-concentration clusters do not shrink but
may even expand, see also the lower two simulations in
Fig.~\ref{fig:evolv}. The top left panel of Figure~\ref{fig:rates} plots
the relative rates versus each other.  While there is a clear tendency
for mass loss to anti-correlate with shrinking, the spread in this
diagram is considerable.

We like to point out that the shrinking rates are very small, of the
order of a few percent per Gyr, too small to protect the cluster from
continued tidal stripping, i.e.\ self-limitation of disk shocks does
not apply here, as is also evident from the undamped mass-loss rates.

%%%%%%%%%%%%%%%%%%%%%%%%%%%%%%%%%%%%%%%%
\ifpreprint\relax\else
  \placefigure{fig:tail:morph}
\fi
%%%%%%%%%%%%%%%%%%%%%%%%%%%%%%%%%%%%%%%%
\subsection{Tidal Tail Morphology}
%%%%%%%%%%%%%%%%%%%%%%%%%%%%%%%%%%%%%%%%%%%%%%%%%%%%%%%%%%%%%%%%%%%%%%%%%%
The stars lost from the cluster form two tidal tails, one leading the
cluster and one trailing it. We will discuss the dynamics of the tidal
tail in detail in the next section. Here, we want to give an impression
of the morphology of these tails.

For the model whose remnant cluster best fits Pal\,5 (see section
\ref{sec:compare:cluster}), hereafter `model A', Figure\
\ref{fig:tail:morph} shows snapshots of (one out of ten) stars in tail and
cluster: the particle distribution projected onto the (instantaneous)
orbital plane at time intervals of 125\,Myr. At the first time shown,
the cluster has not yet been tidally shocked and all the particles are
still inside the cluster, after that, the two tails develop.

There are several notable observations possible from this Figure. First,
near the cluster, the tail shows the typical S-shape, also visible in
the observations of Pal\,5, which originates from the trailing tail
being at larger and the leading tail at smaller distance from the
Galactic center.

Second, the tail morphology varies between thin and long near
perigalacticon (e.g.\ $t=-0.45\,$Gyr), and short and thick near
apogalacticon (e.g.\ $t=-2.075$, $-0.075\,$Gyr), which is due to
variations of the orbital velocity and hence the separation between tail
stars and cluster. This is analogous to cars on a motorway: slow driving
with short separations as in a traffic jam and fast driving with large
distances correspond to the situations near apogalacticon and
perigalacticon, respectively.

Third, the tail neatly huddles against the cluster orbit at most of the
times; only rarely near apogalacticon do some stars of the trailing
(leading) tail appear closer (further). Fourth, occasionally near
apogalacticon, the tidal tail shows a streaky structure. Each of these
streaks corresponds to a swarm of stars that has been set loose by the
same disk shock and fills apparently a streak-like phase-space volume,
which occasionally projects also onto a streak in configuration space
shown in the Figure.
%%%%%%%%%%%%%%%%%%%%%%%%%%%%%%%%%%%%%%%%%%%%%%%%%%%%%%%%%%%%%%%%%%%%%%%%%%
\section{The Dynamics of Tidal Tails} \label{sec:tail}
%%%%%%%%%%%%%%%%%%%%%%%%%%%%%%%%%%%%%%%%%%%%%%%%%%%%%%%%%%%%%%%%%%%%%%%%%%
The gravitational influence of the cluster ceases not abruptly beyond
its tidal radius, but it is completely negligible for stars that have
drifted away from the cluster by more than a few tidal radii. Similarly,
the self-gravity within the tidal tail is unimportant, simply because
the tidal-tail stars move too fast w.r.t.\ each other\footnote{This may
be different in tidal tails emerging from galaxy interactions, where
self-gravity in the tails, possibly in conjunction with gas-dynamics,
may lead to the formation of bound clumps that then become so-called
`tidal dwarf galaxies' \citep[e.g.][]{duc00}.}.  Thus, to very good
approximation, the motion of a tidal tail star is governed by the
gravitational potential $\Phi$ of the Milky Way only.

For a time-invariant and smooth Galactic gravitational field, such as
adopted in the simulations (but see section~\ref{sec:disc:pal5:tail} for
a discussion of deviations from this ideal), a tidal-tail star moves on
a Galactic orbit which is very similar to that of the cluster
itself. The offset in orbital energy may be estimated to be \citep{j98}
\begin{equation} \label{eq:dEorb}
  \delta E_{\mathrm{{orb}}} \simeq
	\pm\;r_{\mathrm{tid}} \; |\boldsymbol{\nabla}\Phi|.
\end{equation}
This follows from the fact that when the star leaves the cluster its
cluster-internal energy approximately vanishes and its velocity is very
similar to that of the cluster, which leaves only the difference in
potential energy as significant contribution to $\delta
E_{\mathrm{{orb}}}$.

In reality, the star may leave the cluster with a higher than just
escape energy, and $|\delta E_{\mathrm{{orb}}}|$ follows a distribution
with a tail to larger values, but in any case $|\delta
E_{\mathrm{{orb}}}|\ll|E_{\mathrm{orb}}|$. Thus, a star in the tidal
tail moves on an orbit with slightly higher (lower) orbital energy than
the cluster, and hence slightly longer (shorter) orbital period. Over
time, the longer (shorter) orbital period results in a lag (lead) of the
star w.r.t.\ the cluster, i.e.\ positive and negative $\delta
E_{\mathrm{{orb}}}$ result in a trailing and leading tidal tail,
respectively. Since the trailing tail moves on a higher orbital energy,
it has a larger radius.

If we assume that a tail star leaves the cluster at a larger or smaller
galactocentric radius, but the same Galactic azimuth and latitude, and
the same velocity as the cluster, then we expect it to have the same
orbital eccentricity as the cluster and to move in the same orbital
plane (paper~II). In reality, we expect deviations from this ideal, but,
of course, the orbital eccentricity and inclination should only slightly
differ from that of the cluster orbit.

\subsection{Coordinates for Tidal-Tail Stars}\label{sec:tail:coords}
%%%%%%%%%%%%%%%%%%%%%%%%%%%%%%%%%%%%%%%%%%%%%%%%%%%%%%%%%%%%%%%%%%%%%%%%%%
Since the orbits of a tidal-tail star and of the cluster are so similar,
the position of a tail star may be referred to the cluster's
trajectory. In paper~II, we gave a simple model for the kinematics of
stars in the tail, which was based on the assumption that its orbit has
the same eccentricity as that of the cluster and differs only in
energy. If we further assume that the Milky Way has a flat rotation
curve, we have
\begin{equation} \label{eq:offset}
  \frac{|\delta\boldsymbol{x}|}{|\boldsymbol{x}_{\mathrm{orb}}|} 
  \approx \frac{|\delta P|}{P} 
\end{equation}
where $\delta P$ is the difference between the orbital period of the
tail star and the period $P$ of the cluster, while $\delta
\boldsymbol{x}\equiv\boldsymbol{x}-\boldsymbol{x}_{\mathrm{orb}}$ is the
positional offset of a tail star from the cluster trajectory at the same
\emph{orbital phase}, not the same time, see Fig.~\ref{fig:tailcoords}
for an illustration.

%%%%%%%%%%%%%%%%%%%%%%%%%%%%%%%%%%%%%%%%
\ifpreprint
  \begin{figure}[t]
    \centerline{\resizebox{47mm}{!}{\includegraphics{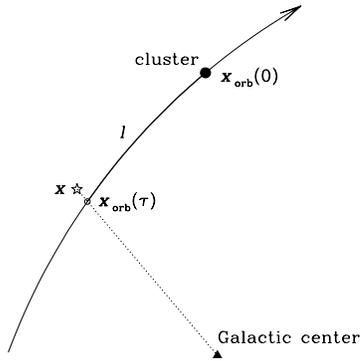}}}
    \figcaption{Illustration for the definition of tail coordinates, see
    section~\ref{sec:tail:coords} for details. \label{fig:tailcoords}}
  \end{figure}
\else
  \placefigure{fig:tailcoords}
\fi
%%%%%%%%%%%%%%%%%%%%%%%%%%%%%%%%%%%%%%%%
Since at the same phase $\delta\boldsymbol{x}\parallel
\boldsymbol{x}_{\mathrm{orb}}$, equation (\ref{eq:offset}) gives a
simple relation between the \emph{perpendicular\/} positional offset of
the tail star from the cluster trajectory and the period difference,
which is directly related to the positional offset \emph{along\/} the
trajectory. In paper~II, we used this simple relation to estimate the
mean drift rate, and consequently the mass-loss rate, from the mean
radial orbital offset of the tidal tail of Pal\,5 from its orbit.  For a
flat-rotation-curve Galaxy, we further have (for $\delta P\ll P$)
\begin{equation} \label{eq:dP}
  \frac{\delta P}{P} 
  = \exp\left(\frac{\delta E_{\mathrm{{orb}}}}{v_{\mathrm{c}}^2}\right) - 1
  \approx \frac{\delta E_{\mathrm{{orb}}}}{v_{\mathrm{c}}^2}.
\end{equation}

%%%%%%%%%%%%%%%%%%%%%%%%%%%%%%%%%%%%%%%%
\ifpreprint
  \begin{figure*}[t]
    \centerline{\resizebox{130mm}{!}{\includegraphics{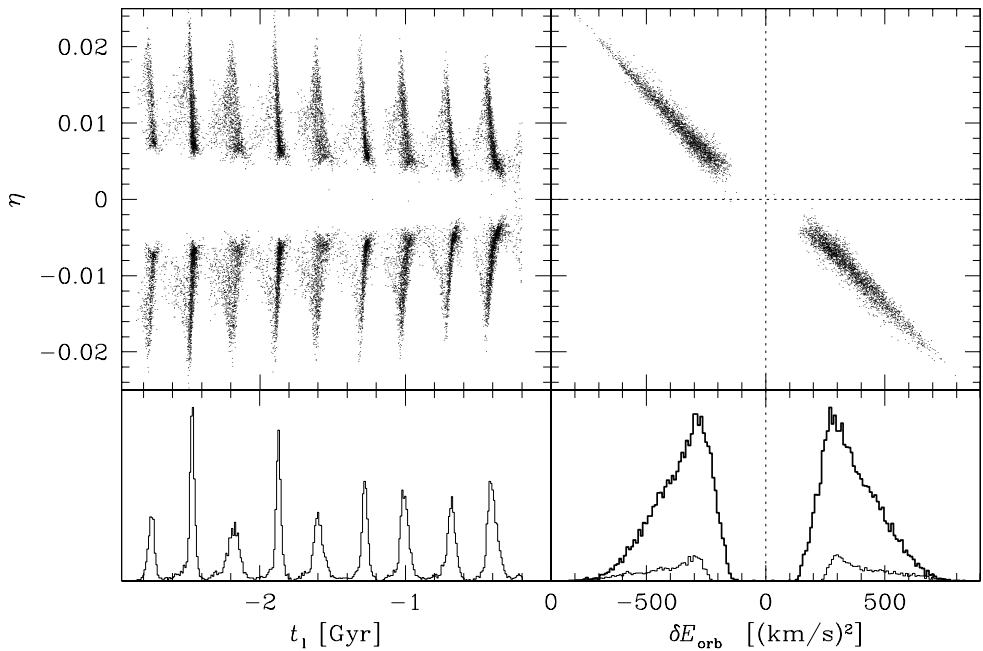}}}
    \figcaption{Distribution of tidal-tail stars over the dimensionless
    drift rate $\eta$, time $t_{\mathrm{l}}$ of loss from the cluster,
    and orbital energy offset $\delta E_{\mathrm{{orb}}}$ from the
    cluster. The values are obtained at $t=0$ from model A
    ($R_{\mathrm{t}}=10\,$kpc, $W_0=2.75$, and
    $M_0=20000\,M_\odot$). Tidal-tail stars with
    $t_{\mathrm{l}}>-0.2\,$Gyr have been omitted. The thin lower
    histogram in the bottom right panel is for stars with
    $t_{\mathrm{l}}\in[-2.5,-2.4]\,$Gyr. In the upper right panel only
    one out of four bodies is plotted. \label{fig:eta}}
  \end{figure*}
\fi
%%%%%%%%%%%%%%%%%%%%%%%%%%%%%%%%%%%%%%%%
Motivated by this model, we now introduce simple coordinates for stars
in the tidal-tail. For a tail star at position $\boldsymbol{x}$, we
first find the reference point $\boldsymbol{x}_{\mathrm{orb}}$ on the
local cluster orbit by minimizing $|\boldsymbol{x}\times
\boldsymbol{x}_{\mathrm{orb}}(\tau)|$ w.r.t.\ the time parameter $\tau$
along the cluster orbit, see Figure~\ref{fig:tailcoords} for an
illustration. With $\tau$ defined such that
$\boldsymbol{x}_{\mathrm{orb}}(0) =\boldsymbol{x}_0$, we can use the
time $\tau$ directly as the tail star's (relative) coordinate along the
tail. For the trailing (sketched in Fig.~\ref{fig:tailcoords}) and
leading stars, $\tau<0$ and $\tau>0$, respectively. Note that, in
contrast to the true positional offset between cluster and tail stars,
$\tau$ is not subject to `seasonal' variations between apogalacticon and
perigalacticon seen in Fig.~\ref{fig:tail:morph}, since these are caused
by variations in orbital velocity.

As further coordinates we use the components
\begin{mathletters} \label{eq:ss}
  \begin{eqnarray}
    \delta x_\parallel &\equiv&
    \delta\boldsymbol{x} \cdot \boldsymbol{x}_{\mathrm{orb}} /
	 |\boldsymbol{x}_{\mathrm{orb}}|, \\
    \delta x_\perp     &\equiv&
    \delta\boldsymbol{x} \cdot \boldsymbol{L}_{\mathrm{orb}} /
			 |\boldsymbol{L}_{\mathrm{orb}}|,
  \end{eqnarray}
\end{mathletters}
of $\delta\boldsymbol{x}$ parallel to $\boldsymbol{x}_{\mathrm{orb}}$ (in
the orbital plane) and parallel to the angular momentum vector
$\boldsymbol{L}_{\mathrm{orb}}\equiv\boldsymbol{x}_{\mathrm{orb}}\times
\dot{\boldsymbol{x}}_{\mathrm{orb}}$ (perpendicular to the orbital plane and
the paper in Figs.~\ref{fig:tail:morph} and \ref{fig:tailcoords}).

We may go a step further and define the dimensionless coordinates
\begin{mathletters} \label{eq:eta}
  \begin{eqnarray}
    \eta            &\equiv& \tau/(t-t_{\mathrm{l}}),                   \\
    \zeta_\parallel &\equiv& \delta x_\parallel / |\boldsymbol{x}_{\mathrm{orb}}|,\\
    \zeta_\perp     &\equiv& \delta x_\perp     / |\boldsymbol{x}_{\mathrm{orb}}|,
  \end{eqnarray}
\end{mathletters}
where $t_{\mathrm{l}}$ is the time when the star was lost from the
cluster. $\eta$ essentially is a \emph{dimensionless drift rate}, while
$\zeta_\parallel$ and $\zeta_\perp$ are \emph{dimensionless offsets\/} in
and out of the orbital plane.

The time $\tau$ corresponds to a phase difference
$\delta\theta=\tau\omega$, which originates from the frequency
difference $\delta\omega$ acting over the time $t-t_{\mathrm{l}}$, i.e.\
$\delta\theta=(t-t_{\mathrm{l}})\delta\omega$. We thus have
$\eta=\delta\omega/\omega=-\delta P/P$ and equations (\ref{eq:offset}) and
(\ref{eq:dP}) yield
\begin{mathletters} \label{eq:ede}
  \begin{eqnarray}
   \zeta_\parallel &\approx&-\eta			\label{eq:zeta} \\
  -\eta            &\approx& \delta E_{\mathrm{{orb}}} /
			v_{\mathrm{c}}^2 		\label{eq:etadE}
  \end{eqnarray}
\end{mathletters}
and $\zeta_\perp\approx0$. That is, not only should these coordinates be
conserved, i.e.\ all the kinematics is contained in their construction,
but they are related in a very simple way. However, the model on which
equation~(\ref{eq:offset}) is based ignores differences in orbital
eccentricity and inclination between tail-star and cluster. Since the
orbital period is mainly a function of orbital energy, whereas the
orbital position at fixed phase also depends sensitively on the
eccentricity and inclination, we expect deviations from our ideal to
mainly affect (\ref{eq:zeta}) and $\zeta_\perp\approx0$, but not so much
(\ref{eq:etadE}).

\subsection{Kinematics of the Tidal Tail}
%%%%%%%%%%%%%%%%%%%%%%%%%%%%%%%%%%%%%%%%%%%%%%%%%%%%%%%%%%%%%%%%%%%%%%%%%%
We are now going to investigate in some detail the distribution of
tidal-tail stars over the relative coordinates $(\eta,
\,\zeta_\parallel, \,\zeta_\perp)$, $\delta E_{\mathrm{{orb}}}$, and
$t_{\mathrm{l}}$, thereby also testing the simple relations
(\ref{eq:ede}) above.  For this purpose, we concentrate on the one
simulation already used in Fig.~\ref{fig:tail:morph} above, which is one
that gives a good description of Pal\,5 (see
Section~\ref{sec:compare}). In this model the cluster originally had
$r_{\mathrm{lim}}=53\,$pc.

\ifpreprint \relax \else
  \placefigure{fig:eta}
\fi
In the top left panel of Figure~\ref{fig:eta}, we plot the dimensionless
drift rate $\eta$, computed at the end of the simulation, vs.\ the time
$t_{\mathrm{l}}$ of loss from the cluster (evaluated as the last instant
when the star crossed $r_{\mathrm{lim}}$ outwards). Clearly, each disk
shock triggers the loss of a swarm of stars with a spread in $\eta$
which is almost symmetric for trailing and leading tail. The morphology
in ($\eta,t_{\mathrm{l}}$) of each of these shocks is the same. First
the stars with largest $|\eta|$ escape, because they have highest
velocity, and somewhat later those with smaller $|\eta|$.  The smallest
$|\eta|$ of any tail star decreases towards later times, because the
cluster loses mass and escaping from it becomes easier.

The drift rates $\eta$ follow a broad distribution with a factor of four
between minimum and maximum. Therefore, a swarm of tail stars set loose
by one disk shock quickly disperses along the tail. In particular, the
fastest drifting tail stars will soon run into the slowest drifting ones
lost at the previous shock. This implies that at any distance from the
cluster the tail is a superposition of stars lost at various epochs,
illustrated also in Fig.~\ref{fig:taildens} below. We may estimate the
number of shocks contributing to the tidal tail at some `distance'
$\tau$ by dividing the difference between maximum and minimum
$t-t_{\mathrm{l}}$ by the average time $t_{\mathrm{disk}}$ between shocks
\begin{equation}
  n \approx (\eta_{\mathrm{min}}^{-1}-\eta_{\mathrm{max}}^{-1})
  \frac{\tau}{t_{\mathrm{disk}}}
  \sim  \frac{3}{4\eta_{\mathrm{min}}} \frac{\tau}{t_{\mathrm{disk}}}.
\end{equation}
With $\eta_{\mathrm{min}}\sim 0.005$ and
$t_{\mathrm{disk}}\approx300\,$Myr, this gives $n\sim0.5(\tau/$Myr). For
the current situation of Pal\,5, this translates into $n\sim 2(l/$deg)
with angular separation $l$ from the cluster (assuming
$v_{\mathrm{orb}}\sim100\,$km\,s$^{-1}$ for the cluster's orbital
velocity).

The maximum drift rate is still rather low: about 2 percent, i.e.\ it
would take $\sim50$ orbits $\sim15\,$Gyr for the fastest first escapees
to cover a full orbital period, i.e.\ to fill a great circle on the sky.

The top-right panel of Fig.~\ref{fig:eta} shows that there is indeed a
tight relation between $\delta E_{\mathrm{{orb}}}$ and $\eta$, as
expected from our equation (\ref{eq:etadE}). However, there is a
considerable spread, in particular at small $|\eta|$, which persists if
we restrict the analysis to larger $|t_{\mathrm{l}}|$ (excluding stars
still in the neighbourhood of the cluster). There is also a systematic
deviation in the sense that the linear relations for the trailing and
leading tail have non-vanishing zero points.

The bottom panels of Fig.~\ref{fig:eta} show histograms of
$t_{\mathrm{l}}$, i.e.\ the mass-loss rate, and $\delta
E_{\mathrm{{orb}}}$.  A comparison with Fig.~\ref{fig:tfield} shows that
the amount of mass lost in a shock correlates with the strength of the
tidal field, as expected.  The distribution in $\delta
E_{\mathrm{{orb}}}$ is, of course, a superposition of the distributions
created by each of the disk shocks and the thin histogram shows the
distribution from the single shock at $-2.45\,$Gyr. The fact that the
inner edge of the distribution (towards small $|\delta
E_{\mathrm{{orb}}}|$) is not abrupt is mainly due to the fact that the
cluster weakened with time, allowing smaller $|\delta
E_{\mathrm{{orb}}}|$ at later epochs.  The typical scale for $\delta
E_{\mathrm{{orb}}}$ is $\sim300\,$(km\,s${-1}$)$2$. This value indeed
follows from equation (\ref{eq:dEorb}) for a cluster with
$r_{\mathrm{tid}}\sim60\,$pc at $R\sim10\,$kpc in a logarithmic
potential with $v_{\mathrm{c}}\sim220\,$km\,s$^{-1}$. The distribution
in $\delta E_{\mathrm{{orb}}}$ is not quite, but still remarkably
symmetric.

%%%%%%%%%%%%%%%%%%%%%%%%%%%%%%%%%%%%%%%%
\ifpreprint
  \begin{figure}[t]
    \centerline{\resizebox{75mm}{!}{\includegraphics{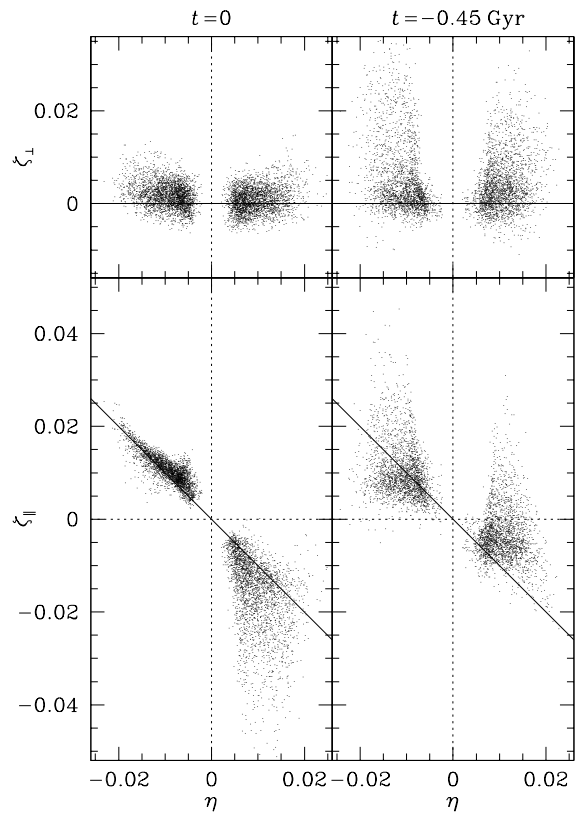}}}
    \figcaption{Plots of $\zeta_\perp$ (\emph{top}) and
    $\zeta_\parallel$ (\emph{bottom}) vs.\ $\eta$ for model A (the same
    as in Fig.~\ref{fig:eta}) at $t=0$ (\emph{left}, near apocenter)
    and $t=-0.45\,$Gyr (\emph{right}, near apocenter). The thin lines
    indicate the relations (\ref{eq:zeta}) and $\zeta_\perp=0$, which
    are expected from naive modeling. For clarity, only one out of four
    bodies is plotted. \label{fig:zeta}}
  \end{figure}
\else
  \placefigure{fig:zeta}
\fi
%%%%%%%%%%%%%%%%%%%%%%%%%%%%%%%%%%%%%%%%
In Figure~\ref{fig:zeta}, we plot the dimensionless orbital offsets
$\zeta_\parallel$ and $\zeta_\perp$ vs.\ $\eta$ computed at two
different epochs, once near apogalacticon and once near perigalacticon.
Quite obviously, the relation $\eta\approx-\zeta_\parallel$
(\ref{eq:zeta}) does not generally hold, only for the trailing tail at
$t=0$ we find reasonable agreement. We also find that $\zeta_\perp$
deviates from zero substantially in the sense that typical values are
significant compared to $\eta$. A comparison between the two epochs
clearly shows that $\zeta_\parallel$ and $\zeta_\perp$ are in fact
\emph{not\/} conserved. This, together with the violation of
$\zeta_\parallel\approx-\eta$ and $\zeta_\perp\approx0$ implies, in
accordance with the discussion following equations (\ref{eq:ede}), that
the orbital eccentricities and inclinations of tail stars differ
significantly and systematically from that of the cluster.  On the other
hand, $\delta E_{\mathrm{{orb}}}$ shows only marginal changes with time
(not shown in a Figure), while $\eta$ changes significantly with time,
but still much less than 100\%, in contrast to $\zeta_\parallel$ and
$\zeta_\perp$.

We may thus conclude that in practice the dimensionless drift rate
$\eta$ is reasonably well conserved and correlates well with $\delta
E_{\mathrm{{orb}}}$, while $\zeta_\parallel$ and $\zeta_\perp$ are
subject to substantial time evolution.  These conclusions are supported
by all simulations, not just the one presented in Figures~\ref{fig:eta}
and \ref{fig:zeta}.

%%%%%%%%%%%%%%%%%%%%%%%%%%%%%%%%%%%%%%%%
\ifpreprint
  \begin{figure}[t]
    \centerline{\resizebox{85mm}{!}{\includegraphics{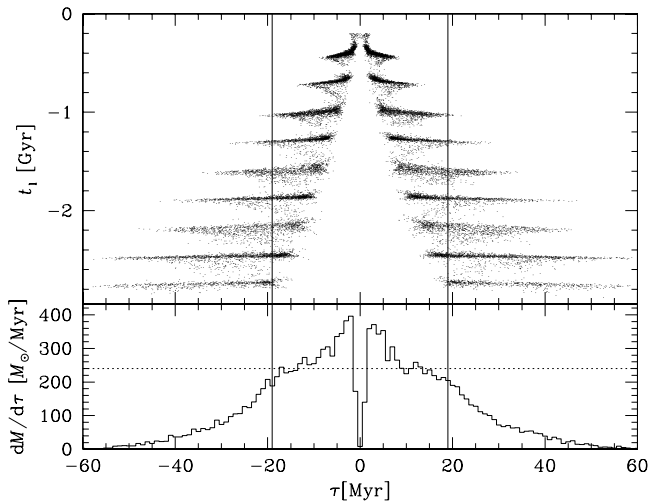}}}
    \figcaption{Distribution of tidal-tail stars in $\tau$ and time
    $t_{\mathrm{l}}$ of loss from the cluster for model A (the same as in
    Fig.~\ref{fig:eta}). The two vertical lines bracket the range in
    $\tau$ inaccessible to stars with $t_{\mathrm{l}}<-2.95\,$Gyr. That
    is, stars lost prior to the start of the simulation, which are, of
    course, not contained in our model, contribute only to
    $|\tau|>19\,$Myr. \label{fig:taildens}}
  \end{figure}
\fi
%%%%%%%%%%%%%%%%%%%%%%%%%%%%%%%%%%%%%%%%
\subsection{Stellar Density along the Tail} \label{sec:dens_tail}
%%%%%%%%%%%%%%%%%%%%%%%%%%%%%%%%%%%%%%%%%%%%%%%%%%%%%%%%%%%%%%%%%%%%%%%%%%
The linear stellar density $\varrho=\mathrm{d}M/\mathrm{d}s$ along the
tail, or, similarly, the circularly averaged surface density
$\Sigma(R)$ of stars (tail \& cluster) are easily observable from Pal\,5
(see Fig.~\ref{fig:surf}, paper~II). A study by \cite{jsh99} of the
formation of tidal tails due to Galactic tides acting on globular
clusters or satellite galaxies suggest that $\Sigma\propto R^{-1}$, or
equivalently $\varrho\approx\mathrm{const}$\footnote{In another study
\cite{jcg02} reported much steeper profiles ($\Sigma\propto R^{-3}$).
However, these occurred in much more massive satellites and over larger
parts of the orbit and are not relevant to the situation studied here.}

We may identify three factors that determine $\varrho(s)$: the mass-loss
rate $\dot{M}$; the distribution $f(\eta)$ of dimensionless drift rates,
which may be a function of the time $t_{\mathrm{l}}$ of loss from the
cluster; and the orbital kinematics that map the phase-distance $\tau$
into physical distance $s$ along the orbit. The first two factors, which
depend mainly on the general properties of the orbit (period and
perigalactic radius) and the cluster, determine the distribution of tail
stars over $\tau$
\begin{equation} \label{eq:dMdtau}
  \frac{\mathrm{d}M{}}{\mathrm{d}\tau} = 
  \int_{-\infty}^t\! \mathrm{d} t_{\mathrm{l}}\; \dot{M}(t_{\mathrm{l}})
  \int_{-\infty}^\infty \!\mathrm{d} \eta\; f(\eta; t_{\mathrm{l}})\;
  \delta\big(\tau-\eta(t-t_{\mathrm{l}})\big).
\end{equation}
The last factor depends on the current position on the orbit and yields
\begin{equation}
  \varrho(s;t) = \frac{1}{|\boldsymbol{v}_{\mathrm{orb}}(\tau)|}
  \frac{\mathrm{d}M{}}{\mathrm{d}\tau}
\end{equation}
where $s=\int_0^\tau |\boldsymbol{v}_{\mathrm{orb}}(\tau^\prime)|\,
\mathrm{d}\tau^\prime$ calculated at the current time $t$.

A constant $\varrho(s)$, such as found in the simulations by Johnston et
al., emerges from a constant (orbit-averaged) $\dot{M}$ in conjunction
with a time-invariant distribution $f(\eta)$ of drift rates and
$|\boldsymbol{v}_{\mathrm{orb}}|\approx\mathrm{const}$ over the extent of
the tail.  This latter condition is not satisfied whenever the tidal
tail spans a significant range in phases on an eccentric orbit.

\ifpreprint \relax \else
  \placefigure{fig:taildens}
\fi
As we saw in Fig.~\ref{fig:eta}, the distribution over drift rates
$\eta$ is actually \emph{not\/} time invariant for our models: the drift
rates decrease, because the cluster is significantly weakened by the
continued loss of stars.  With a constant (orbit averaged) mass-loss
rate, this results in $\mathrm{d}M/\mathrm{d}\tau$ slightly decreasing
with $|\tau|$. This is illustrated in Figure~\ref{fig:taildens}, which
plots the distribution of tail stars over $\tau$ and
$t_{\mathrm{l}}$. In the top panel, it is evident that the typical
$|\tau|$ increases slightly faster than linear with $t-t_{\mathrm{l}}$
(the inner envelopes are not straight lines). The plot of
$\mathrm{d}M/\mathrm{d}\tau$ in the bottom panel shows (i) some
`bumpiness' at the $\lesssim20\%$ level and (ii) a slight decrease with
$|\tau|$ (data for $|\tau|>19\,$Myr are incomplete since tail stars with
$t_{\mathrm{l}}<-2.95\,$Gyr are lacking from our models, see also the
discussion at the end of section \ref{sec:simul:init}). For the linear
density $\varrho$, the latter implies a slight decrease with $s$ and a
steeper than $R^{-1}$ decrease for $\Sigma$.  Thus, we expect
non-constant $\varrho(s)$ even for a (orbit-averaged) constant mass-loss
rate and a tail spanning a small range of
$|\boldsymbol{v}_{\mathrm{orb}}|$,
\emph{if\/} the mass loss was substantial.

For clusters or satellites which undergo only non-substantial mass loss,
the distribution over $\eta$ should indeed be independent of time and
consequently $\mathrm{d}M/\mathrm{d}\tau\approx\mathrm{const}$,
consistent with the results of the aforementioned study by Johnston et
al..

The bumpiness of $\mathrm{d}M/\mathrm{d}\tau$, which of course
translates in a bumpiness of $\varrho(s)$, is caused by individual
shocks and the fact that for small $\tau$ only few of them contribute.
These bumps are expected to be symmetric, i.e.\ occur in trailing and
leading tail at about the same distance from the cluster. For larger
$\tau$ or distances $s$ from the cluster these inhomogeneities are
averaged out by the superposition of many swarms of stars lost from the
cluster at different shock events.

\subsection{Drift Rates and Tail Kinematics of the Models}
%%%%%%%%%%%%%%%%%%%%%%%%%%%%%%%%%%%%%%%%%%%%%%%%%%%%%%%%%%%%%%%%%%%%%%%%%%
In order to investigate the dependence of the tail properties on the
model parameters, we computed for each model of a cluster that survived
until today the mean and dispersion of $\eta$, $\zeta_\parallel$,
$\zeta_\perp$, and $\delta E_{\mathrm{{orb}}}$ evaluated at $t=0$ for
both the leading and trailing tail.

%%%%%%%%%%%%%%%%%%%%%%%%%%%%%%%%%%%%%%%%
\ifpreprint
  \begin{figure}[t]
    \centerline{\resizebox{84mm}{!}{\includegraphics{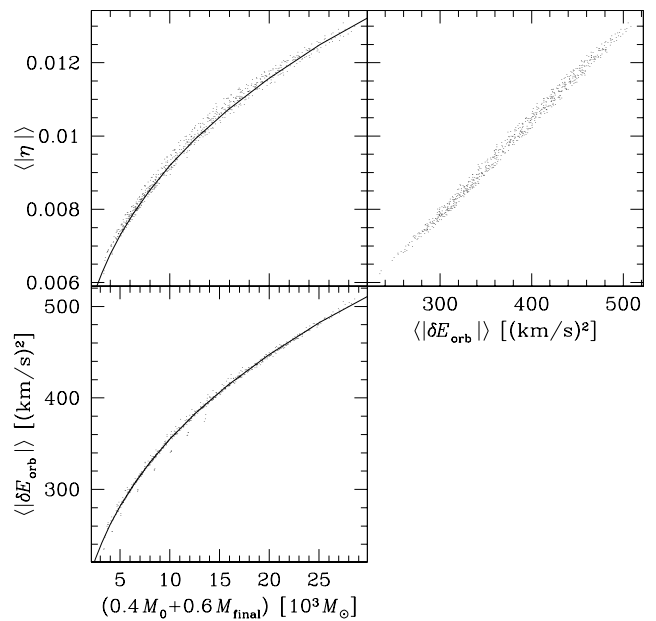}}}
    \figcaption{For all 695 simulations with a cluster surviving until
    $t=0$, we plot against each other: the average offset in orbital
    energy $\delta E_{\mathrm{{orb}}}$ between tail stars and cluster;
    the average absolute dimensionless drift rate $\eta$; and the
    weighted mean of the initial and final mass of the cluster. $\delta
    E_{\mathrm{{orb}}}$ and $\eta$ have been calculated at $t=0$. The
    thin curves in the two left panels follow a power-law with exponent
    $1/3$. \label{fig:dEorb}}
  \end{figure}
\else
  \placefigure{fig:dEorb}
\fi
%%%%%%%%%%%%%%%%%%%%%%%%%%%%%%%%%%%%%%%%
The estimate (\ref{eq:dEorb}) for $\delta E_{\mathrm{{orb}}}$ suggests
that $\langle|\delta E_{\mathrm{{orb}}}|\rangle$ depends only on the
mass $M$ of the cluster at the instant of the star leaving it. Because
our models are subject to substantial mass loss, their mass is not
conserved. However, as the bottom-left panel of Figure~\ref{fig:dEorb}
demonstrates, the average $\langle|\delta E_{\mathrm{{orb}}}|\rangle$
over all tail stars (lost over the whole integration interval) is
strongly correlated with some time-averaged cluster mass. In fact
$\langle|\delta E_{\mathrm{{orb}}}|\rangle \propto M^{1/3}$, exactly as
equations~(\ref{eq:dEorb}) and (\ref{eq:rtid}) predict.  The constant of
proportionality in this relation depends on the cluster orbit and the
Galactic potential only, but not on the cluster's intrinsic properties.
The remaining two panels of Fig.~\ref{fig:dEorb} show that the mean
dimensional drift rate $\langle|\eta|\rangle$ is proportional to
$M^{1/3}$, too, and to $\langle|\delta E_{\mathrm{{orb}}}|\rangle$. The
constant of proportionality in this latter relation is
$(196\,$km/s)$^{-2}$, corresponding approximately to
$v_{\mathrm{c}}^{-2}$, in accordance with equation (\ref{eq:etadE}).

With this mass dependence, we obtain for the average density of tail
stars over $\tau$ from equation (\ref{eq:dMdtau})
\begin{equation}
  \left\langle\frac{\mathrm{d}M}{\mathrm{d}\tau}\right\rangle \propto
  \frac{M}{\langle\eta\rangle}
  \propto M^{2/3}
\end{equation}
with $M$ denoting some measure of the cluster mass (final or time
averaged). The constant of proportionality depends on the relative
mass-loss rate $\dot{M}/M_0$ and hence is a function of $R_{\mathrm{t}}$
and $W_0$, but hardly of $M_0$.  Thus, at fixed $(R_{\mathrm{t}},W_0)$,
a smaller cluster mass results in a larger fraction of tail stars at
fixed $\tau$, i.e.\ phase offset from the cluster.

%%%%%%%%%%%%%%%%%%%%%%%%%%%%%%%%%%%%%%%%%%%%%%%%%%%%%%%%%%%%%%%%%%%%%%%%%%
\section{Comparison with Pal\,5} \label{sec:compare}
%%%%%%%%%%%%%%%%%%%%%%%%%%%%%%%%%%%%%%%%%%%%%%%%%%%%%%%%%%%%%%%%%%%%%%%%%%
In this section we compare the 695 models with a surviving cluster to
the observed properties of Pal\,5 and its tail. The objective is to
investigate whether these data are at all consistent with our models and
if so, with which type of models.

%%%%%%%%%%%%%%%%%%%%%%%%%%%%%%%%%%%%%%%%
\ifpreprint
  \begin{figure}[t]
    \centerline{\resizebox{75mm}{!}{\includegraphics{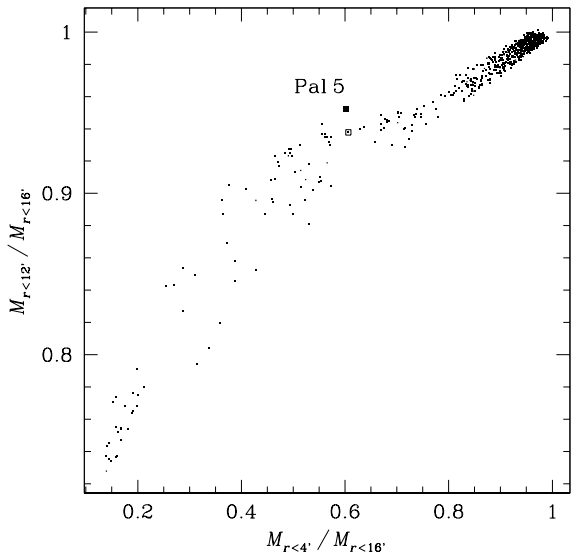}}}
    \figcaption{Distribution of the 695 simulated clusters over the
    fractions of stars within projected $16\arcmin$ that are also within
    projected $4\arcmin$ (\emph{x-axis}) and $12\arcmin$
    (\emph{y-axis}). Note that $16\arcmin$ is the limiting radius of
    Pal\,5 itself, whose data are shown as solid square. The open square
    marks the model which best fits the data for Pal\,5 and is denoted
    `model A' in the text. \label{fig:mfrac}}
  \end{figure}
\else
  \placefigure{fig:mfrac}
\fi
%%%%%%%%%%%%%%%%%%%%%%%%%%%%%%%%%%%%%%%%
\subsection{The Cluster} \label{sec:compare:cluster}
%%%%%%%%%%%%%%%%%%%%%%%%%%%%%%%%%%%%%%%%%%%%%%%%%%%%%%%%%%%%%%%%%%%%%%%%%%
The most astonishing property of Pal\,5, apart from the tails, is its
size, which is about four times larger than expected for a cluster of
its mass and on its orbit (section~\ref{sec:extent}). Instead of
comparing 695 different projected surface density profiles, we plot in
Figure~\ref{fig:mfrac} the fractions of the projected cluster mass
within $16\arcmin$ (the limiting radius of Pal\,5), that are also within
the projected radii of $4\arcmin$ and $12\arcmin$. Most of our models
are in the upper-right corner of the figure: they have nearly $100\%$
for both these fractions, meaning that most of their mass is contained
within $4\arcmin$, the tidal radius of Pal\,5 at perigalacticon.  Thus,
these simulated clusters have survived until today, because they are, in
contrast to Pal\,5, smaller than their tidal radii.

There is, however, also a long tail of models with lower mass fractions
within $4\arcmin$ and $12\arcmin$, corresponding to larger extents. Some
of these actually come close to Pal\,5 in the sense that they have
$\sim95\%$ of their mass within $12\arcmin$ and $\sim60\%$ within
$4\arcmin$. These models have initial parameters $W_0\simeq2-3$,
$R_{\mathrm{t}}\simeq(4.5+2W_0)\,$kpc, and $M_0\simeq20000\,M_\odot$,
while model clusters with $R_{\mathrm{t}}$ larger than that (at fixed
$W_0$) get destroyed and those with lower mass are even more diffuse or
have been disrupted as well.  Initially more concentrated models with
$W_0\ga3$ and also $R_{\mathrm{t}}\simeq(4.5+2W_0)\,$kpc form the tail
in Fig.~\ref{fig:mfrac} towards smaller mass fractions; they have
massive tidal tails but their clusters are not as extended as Pal\,5.

The model that best matches Pal\,5 in Fig.~\ref{fig:mfrac} is that with
initial parameters $R_{\mathrm{t}}=10\,$kpc, $W_0=2.75$, and
$M_0=20000\,M_\odot$; it is marked with an open square and will be
denoted `model A'. This model already featured in Figures
\ref{fig:tail:morph}, \ref{fig:eta},
\ref{fig:zeta}, and \ref{fig:taildens} and will be used in the remainder
of the paper for illustrative purposes.

%%%%%%%%%%%%%%%%%%%%%%%%%%%%%%%%%%%%%%%%
\ifpreprint
  \begin{figure}[t]
    \centerline{\resizebox{75mm}{!}{\includegraphics{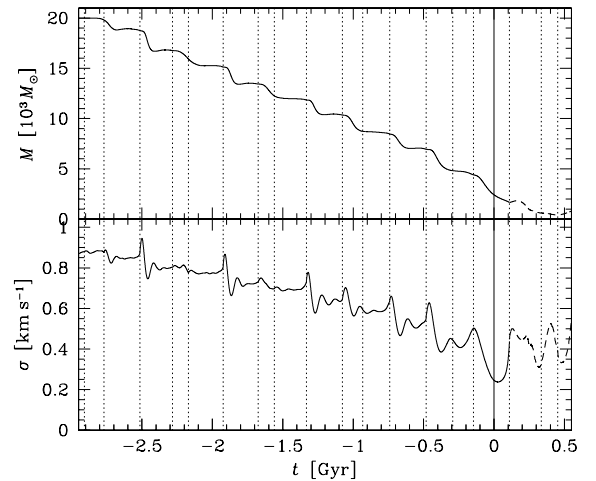}}}
    \figcaption{Time evolution of mass (\emph{top}) and velocity
    dispersion (\emph{bottom}) of model A (with $R_{\mathrm{t}}=10\,$kpc,
    $W_0=2.75$, and $M_0=20000\,M_\odot$). The dotted vertical lines are
    equivalent to those in Fig.~\ref{fig:evolv}. The cluster is
    destroyed by the first future disk crossing at
    $t=110\,$Myr. \label{fig:simulA}}
  \end{figure}
\else
  \placefigure{fig:simulA}
\fi
%%%%%%%%%%%%%%%%%%%%%%%%%%%%%%%%%%%%%%%%
\ifpreprint
  \begin{figure}[t]
    \centerline{\resizebox{75mm}{!}{\includegraphics{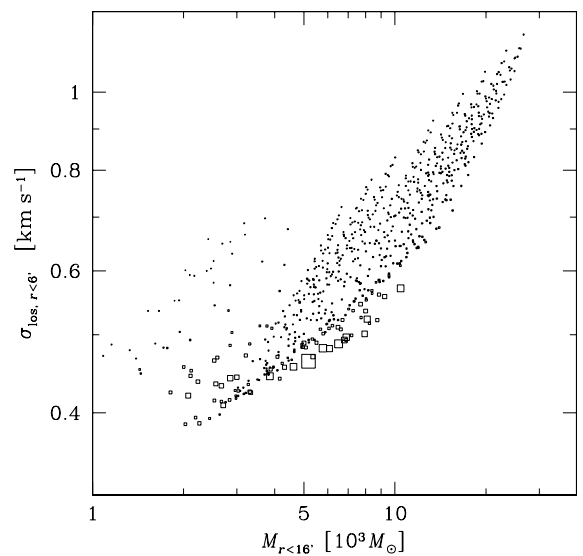}}}
    \figcaption{Distribution of the 695 simulated clusters over the
    final projected mass within 16\arcmin and the line-of-sight velocity
    dispersion within 6\arcmin.  Larger symbol sizes correspond to
    models that are closer to Pal\,5 in their enclosed projected mass
    fractions (Fig.~\ref{fig:mfrac}); the largest symbol refers to model
    A. Note that for Pal\,5, $M_{r<16\arcmin}=5200\pm700M_\odot$ and 
    $\sigma_{\mathrm{los},r<6\arcmin}=0.22^{+0.19}_{-0.10}\,$km\,s$^{-1}$
    (assuming a Gaussian distribution, see also Fig.\ \ref{fig:vrad} and
    the text for a discussion). \label{fig:msig}}
  \end{figure}
\else
  \placefigure{fig:msig}
\fi
%%%%%%%%%%%%%%%%%%%%%%%%%%%%%%%%%%%%%%%%

Figure~\ref{fig:simulA} shows the time evolution of the cluster mass and
velocity dispersion $\sigma$ of model A. All models which come close to
Pal\,5 in their projected mass fractions in Fig.~\ref{fig:mfrac} show
substantial mass loss, which results in a drastic increase in the
cluster's dynamical time $t_{\mathrm{dyn}}$, reflected by the steady
decline of $\sigma$. The increase of $t_{\mathrm{dyn}}$ slows down the
cluster's response to disk shocks: the period of the (damped)
oscillations that follow each shock (in Fig.~\ref{fig:simulA} visible as
oscillations of $\sigma$ subsequent to each dotted vertical line, see
also the discussion in section \ref{sec:simul:evolv:mech})
increases. Eventually, this period approximately equals twice the time
between two shocks, namely those at $t=-146$ and 110\,Myr, such that
just in the middle between these two shocks, i.e.\ \emph{today}, the
cluster is in state of maximal expansion and minimal velocity
dispersion. This property is common to all models with projected density
profile similar to that of Pal\,5.

Thus, all those models that reproduce the overly large extent of Pal\,5
also naturally, and causally connected, predict a very small velocity
dispersion consistent with that actually observed for Pal\,5.

%%%%%%%%%%%%%%%%%%%%%%%%%%%%%%%%%%%%%%%%
\ifpreprint
  \begin{figure}[t]
    \centerline{\resizebox{75mm}{!}{\includegraphics{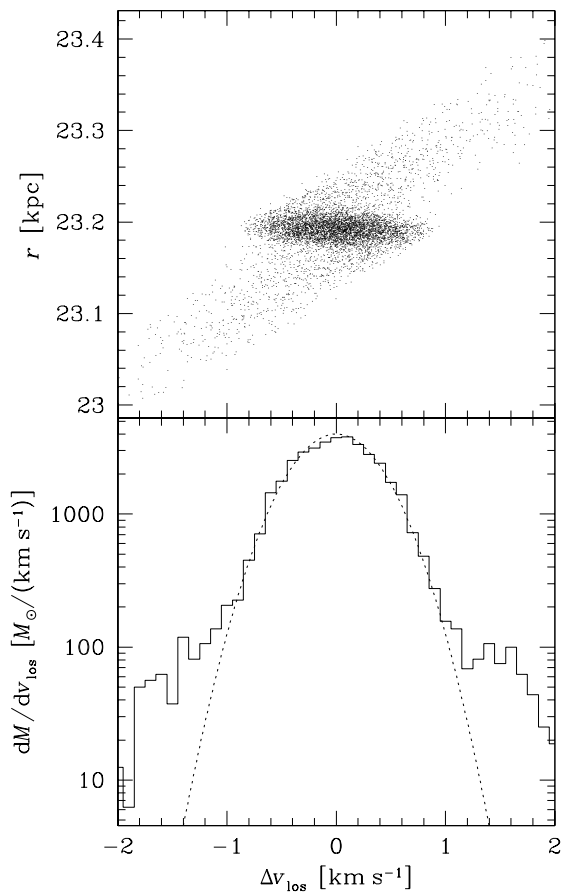}}}
    \figcaption{\emph{Top}: distribution over distance $r$ from the Sun
    and line-of-sight velocity for stars within projected 6\arcmin\ of
    the cluster center for the final state of model A (initial
    parameters: $W_0=2.75$, $R_{\mathrm{t}}=10$, and $M_0=20000M_\odot$).
    \emph{Bottom}: distribution of the same stars over
    $v_{\mathrm{los}}$ and a Gaussian with
    $\sigma_{\mathrm{los}}=0.38\,$km\,s$^{-1}$
    (\emph{dotted}). \label{fig:vrad}}
  \end{figure}
  \begin{figure}[t]
    \centerline{\resizebox{75mm}{!}{\includegraphics{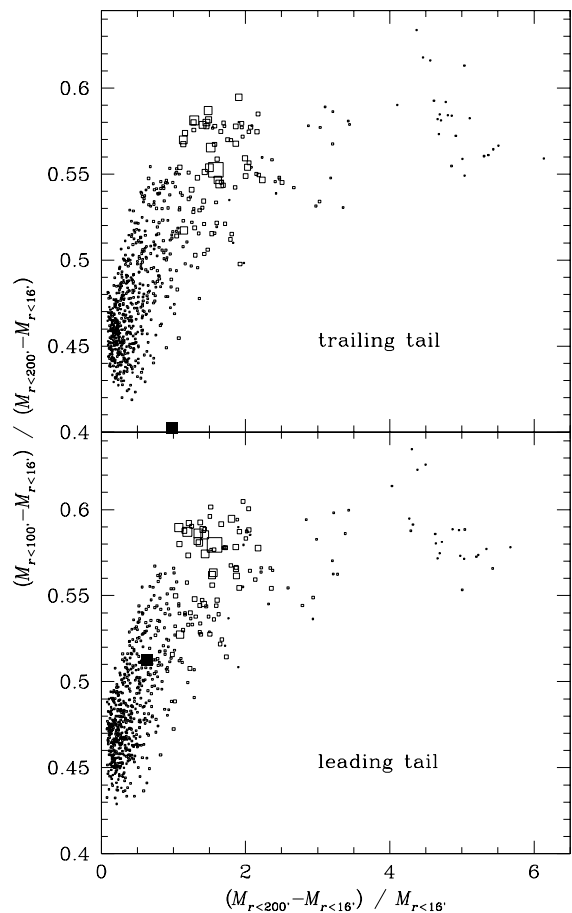}}}
    \figcaption{Distribution of the 695 simulated tails with surviving
    clusters over ($x$-axis) the ratio between mass in the tail out to
    projected $200\arcmin$ and the mass in the cluster (within
    $16\arcmin$) and ($y$-axis) the fraction of tail stars within
    $100\arcmin$ for the trailing (\emph{top}) and leading
    (\emph{bottom}) tail, respectively. The symbol sizes indicate how
    well the model matches the cumulative projected masses of the
    cluster (Fig.~\ref{fig:mfrac}). The solid square represents the
    star-count data for Pal\,5, which have a statistical uncertainty of
    about 0.05 in the $y$ direction. The trailing tail of Pal\,5 has a
    substantial clump at 3 deg from the cluster, compromising the
    measures plotted, in particular that on the $y$ axis.
    \label{fig:mtail}}
  \end{figure}
\fi
%%%%%%%%%%%%%%%%%%%%%%%%%%%%%%%%%%%%%%%%
In Figure~\ref{fig:msig}, we plot for all 695 surviving cluster models
the final mass within projected 16\arcmin\ (the limiting radius of
Pal\,5) and velocity dispersion of stars within projected 6\arcmin. The
models with projected mass profiles most similar to Pal\,5 (shown as
larger symbols) form the lower envelope, i.e.\ have lowest line-of-sight
velocity dispersion $\sigma_{\mathrm{los}}$ at given projected mass. A
direct comparison of these values for $\sigma_{\mathrm{los}}$ with that
determined for Pal\,5 in paper~I (see section \ref{sec:mass_sigma}) is
problematic, as the latter has been obtained under the assumption that
the line-of-sight velocity distribution was Gaussian (high-velocity
tails have been modeled as binary contamination).

%%%%%%%%%%%%%%%%%%%%%%%%%%%%%%%%%%%%%%%%
\ifpreprint\relax\else
  \placefigure{fig:vrad}
\fi
%%%%%%%%%%%%%%%%%%%%%%%%%%%%%%%%%%%%%%%%
In Figure \ref{fig:vrad}, we show for the model that best fits the
projected mass profile (the biggest symbol in Fig.~\ref{fig:msig}) the
distribution of stars within projected 6\arcmin\ of the cluster center
over distance $r$ and line-of-sight velocity $v_{\mathrm{los}}$. The
dotted line in the lower panel represents a Gaussian with
$\sigma_{\mathrm{los}}=0.38\,$km\,s$^{-1}$.  Clearly, the distribution
has non-Gaussian wings, which originate from stars that are just being
lost from the cluster. Thus, the extra-Gaussian tails of the
\emph{observed\/} line-of-sight velocity distribution can at least
partly be explained by cluster dynamics and not entirely by binaries, as
we assumed in paper~I.

The very model shown in Fig.~\ref{fig:vrad} not only gives the best fit
to the \emph{shape\/} of the Pal\,5's mass-profile, but also its
\emph{total mass\/} within 16\arcmin\ of 5177\,$M_\odot$ agrees nicely
with the best estimate for Pal\,5 of $5200\pm700\,M_\odot$ (section\
\ref{sec:mass_sigma}).

\subsection{The Tidal Tail}
%%%%%%%%%%%%%%%%%%%%%%%%%%%%%%%%%%%%%%%%%%%%%%%%%%%%%%%%%%%%%%%%%%%%%%%%%%
\subsubsection{The Radial Profile}
%%%%%%%%%%%%%%%%%%%%%%%%%%%%%%%%%%%%%%%%%%%%%%%%%%%%%%%%%%%%%%%%%%%%%%%%%%
\ifpreprint\relax\else
  \placefigure{fig:mtail}
\fi
%%%%%%%%%%%%%%%%%%%%%%%%%%%%%%%%%%%%%%%%
\ifpreprint
  \begin{figure*}[t]
    \centerline{\resizebox{130mm}{!}{\includegraphics{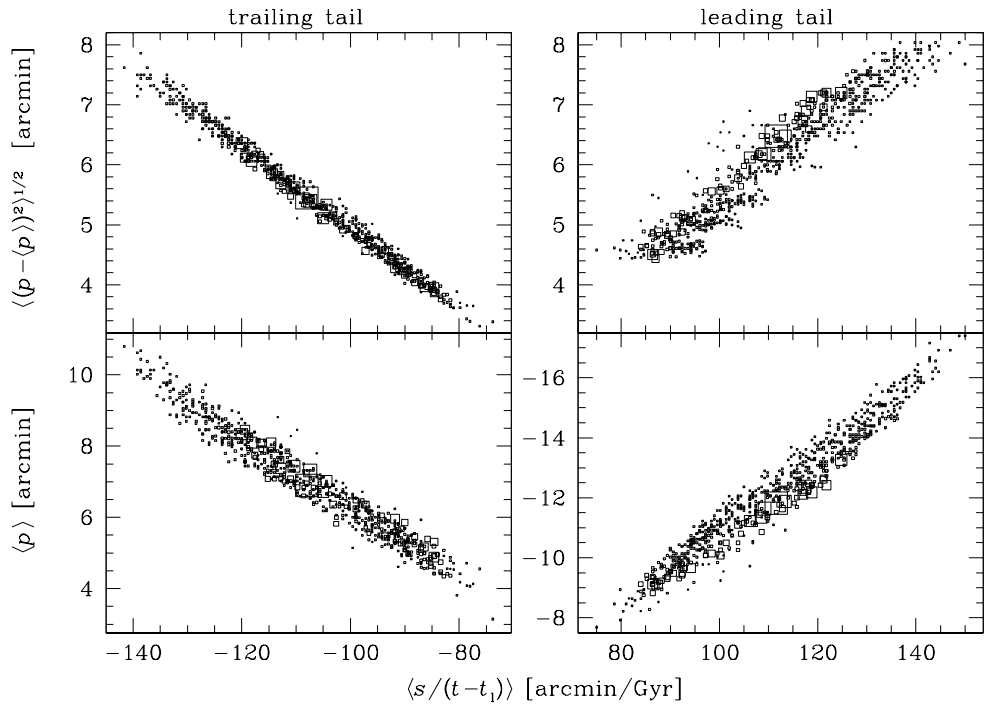}}}
    \figcaption{Distribution of the 695 simulated tails with surviving
    clusters over ($x$-axis) the mean projected drift rate along the
    orbits and ($y$-axes) the mean (\emph{bottom}) and dispersion
    (\emph{top}) of the rectangular displacement $p$ of the tail stars
    from the orbit. Only tail stars with distance (projected arc-length
    $s$ along the tail) from the cluster within
    $[24\arcmin,\,200\arcmin]$ have been used. The symbol sizes indicate
    how well the model matches the cumulative projected masses of the
    cluster (Fig.~\ref{fig:mfrac}). See text for a comparison with
    Pal\,5 \label{fig:tailoff}}
  \end{figure*}
\fi
%%%%%%%%%%%%%%%%%%%%%%%%%%%%%%%%%%%%%%%%
In Figure~\ref{fig:mtail}, we compare the gross radial distribution of
stars in the tail of Pal\,5 with those of the 695 models with surviving
clusters.  On the $x$-axis, we plot, for the trailing (\emph{top}) and
leading (\emph{bottom}) tidal tail, the ratio of tail to cluster
mass. Here, we rather arbitrarily truncated the tail at $200\arcmin$ to
ensure that no stars are missing from the simulations due to our limited
integration time (cf. the discussion following
Fig.~\ref{fig:taildens}). On the $y$-axis of Fig.~\ref{fig:mtail}, we
plot the fraction of the tail mass within $100\arcmin$. This fraction is
related to the density run along the tail: for a constant linear
density, equivalent to $\Sigma\propto r^{-\gamma}$ with $\gamma=1$, we
expect $\sim0.46$, while steeper density profiles with $\gamma>1$ yield
larger values.

We first notice a weak correlation between this ratio and the relative
mass in the tail in the sense that (relatively) more massive tails have
higher fractions of stars inside $100\arcmin$, corresponding to larger
$\gamma$. This is, of course, exactly what we expect from our discussion
in \S\ref{sec:dens_tail}.

Regarding Pal\,5, we see that while its leading tail falls into the
locus of models, the trailing tail does not. This can be entirely
attributed to the clump at $\sim3^\circ$ from the cluster (paper~II, see
also Fig.~\ref{fig:surf}). We also notice that those models which best
fit the properties of the cluster itself (indicated by larger symbol
sizes in Fig.~\ref{fig:mtail}), have tails that are more massive by a
factor 1.5 -- 2 and have a steeper radial density profile (but note that
the statistical uncertainty on this latter property of Pal\,5 is
considerable).

We should note here that for Pal\,5, we actually cannot measure mass
ratios, but merely number count ratios, which are identical to the
former only, if the stellar luminosity functions (LFs) in the cluster
and the various parts of the tail are the same. Actually, \cite{ko04}
reported a significant lack of low-mass stars in the cluster as compared
to its tails, which may explain (at least partly) the apparent
discrepancy in the tail-to-cluster mass ratio.

Thus, we conclude that simultaneous fitting of the projected radial
profile of Pal\,5 \emph{and\/} its tails appears to be difficult, while
each property on its own is inside the locus of the models. However,
statistical uncertainties and our lack of knowledge of mass (rather than
number-count) ratios, i.e.\ of the LF, hampers more quantitative
statements to be made.

Another important point is that the clumpiness of the tidal tail of
Pal\,5, in particular the over-density at $3^\circ$ in the trailing tail,
is not reproduced in any of our models. We will discuss in section
\ref{sec:disc:pal5:tail} the possible implication of this mismatch.

%%%%%%%%%%%%%%%%%%%%%%%%%%%%%%%%%%%%%%%%
\ifpreprint
  \begin{figure*}[t]
    \centerline{\resizebox{120mm}{!}{\includegraphics{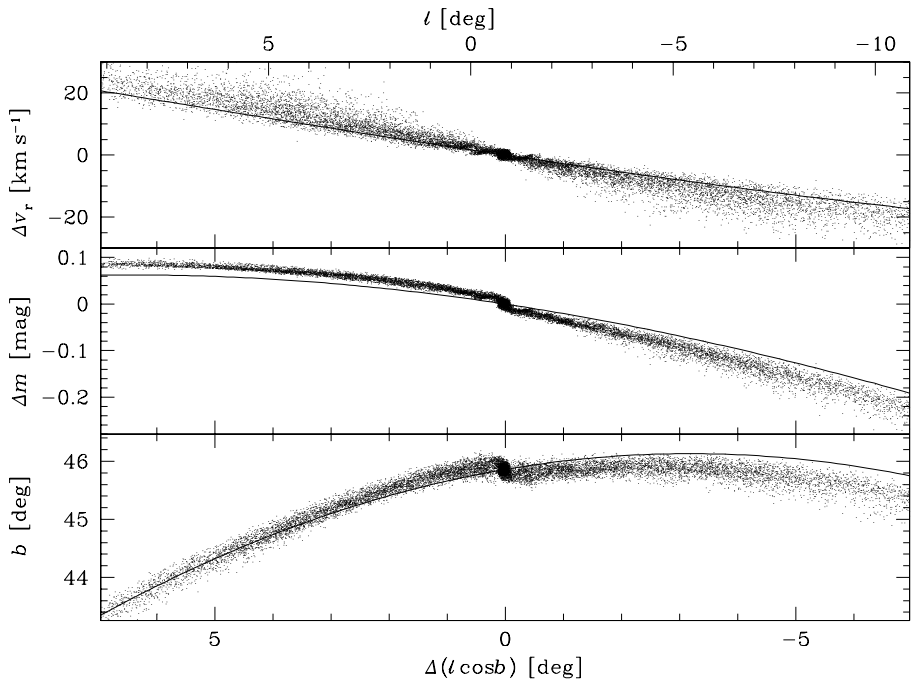}}}
    \figcaption{Projection of model A at $t=0$ (today) onto the sky in
    standard Galactic coordinates (\emph{bottom}) and plots of the
    offset in distance modulus (\emph{middle}) and radial velocity
    (\emph{top}) of the stars from the cluster versus Galactic
    longitude. The thin line represents the orbit. Only every second
    body is plotted. \label{fig:proja}}
  \end{figure*}
\fi
%%%%%%%%%%%%%%%%%%%%%%%%%%%%%%%%%%%%%%%%
\subsubsection{The Offset from the Orbit} \label{sec:offset}
%%%%%%%%%%%%%%%%%%%%%%%%%%%%%%%%%%%%%%%%%%%%%%%%%%%%%%%%%%%%%%%%%%%%%%%%%%
In paper~II, we measured the projected rectangular offset of tail stars
from the projected (assumed) orbit. In order to compare our simulations
to these measurements, we computed the same quantities for our models
(after projection onto the sky as seen from the Sun) as follows. First,
we determined the point on the projected orbit that minimizes the
projected distance to the tail star. We then define $p$ to be the
projected distance with $p>0$ ($p<0$) for tail stars at larger (smaller)
Galactic latitude than the nearest point on the orbit. We define $s$ to
be the arc-length along the projected orbit with $s>0$ and $s<0$ for the
leading and trailing tail, respectively.

%%%%%%%%%%%%%%%%%%%%%%%%%%%%%%%%%%%%%%%%
\ifpreprint\relax\else
  \placefigure{fig:tailoff}
\fi
%%%%%%%%%%%%%%%%%%%%%%%%%%%%%%%%%%%%%%%%
In Figure~\ref{fig:tailoff}, we plot, separately for the trailing
(\emph{left}) and leading (\emph{right}) tails of the all the models
with surviving cluster, the mean (\emph{bottom}) and dispersion
(\emph{top}) of $p$ versus the mean projected drift rate
$s/(t-t_{\mathrm{l}})$.

%%%%%%%%%%%%%%%%%%%%%%%%%%%%%%%%%%%%%%%%
\ifpreprint\relax\else
  \placefigure{fig:proja}
\fi
%%%%%%%%%%%%%%%%%%%%%%%%%%%%%%%%%%%%%%%%
From this Figure, we see that, while the drift rates for the leading and
trailing tail are of comparable size, the mean rectangular distance
$|p|$ is larger by almost a factor of two for the former. This is also
evident from the projection of model A in the bottom panel of
Figure~\ref{fig:proja}. The reason is partial intrinsic, i.e.\ due to
the offsets $\zeta_\parallel$ of the tail stars from the cluster orbit
within the orbital plane (Fig.~\ref{fig:zeta}, left bottom panel), and
partly due to projection effects (because the leading tail is closer to
us and seen under a slightly smaller inclination angle).

In contrast to the modeled tails, the values inferred in paper~II for
the mean rectangular offset of Pal\,5's tail stars are rather similar
for the trailing and leading tail ($11{}^\prime\!.8\pm0{}^\prime\!.5$
and $-10{}^\prime\!.1\pm0{}^\prime\!.8$, respectively). However, this is
essentially a consequence of the implicit assumption, made when adopting
the orbit in paper~II, that the rectangular offsets are similar in size
between the two tails. The fact that the simulations consistently
predict larger $\langle|p|\rangle$ for the leading than the trailing
tail implies that the orbit of Pal\,5 should actually be slightly
different from the one used in paper~II and here; compare, for instance,
the bottom panel of Fig.~\ref{fig:proja} with Fig.~9 of paper~II. (Note,
however, that the uncertainty in the distance to Pal\,5 still dominates
the uncertainty in the orbit).

However, we may still compare the tails' perpendicular structure to the
models by means of the dispersion in $p$. In paper~II, we fitted a
Gaussian to the distributions in $p$ for the trailing and leading tails
of Pal\,5. When assuming that the distribution is actually Gaussian, we
obtain for the dispersions $7{}^\prime\!.8\pm0{}^\prime\!.5$ and
$7{}^\prime\!.3\pm0{}^\prime\!.8$ for the trailing and leading tail,
respectively. Comparing these values with those in
Fig.~\ref{fig:tailoff}, it appears that the tail of Pal\,5 is at the
upper end of the distribution. In particular, these values are somewhat
larger than those for the models that best fit the cluster's cumulative
mass distribution (big symbols in Fig.~\ref{fig:tailoff}).

The mean drift rate estimated for the trailing and leading tail in
paper~II was $\sim180\,$arcmin/Gyr, which is substantially more than
predicted by the models, even by those with as high a dispersion in
$p$ as observed for the tails of Pal\,5. That estimate, however, was
based on rather simplifying assumptions on the relation with $p$ and,
given the model results in Fig.~\ref{fig:tailoff}, a lower value for the
mean drift rate seems more appropriate. In face of the uncertainties, we
cannot make precise statements, except for saying that most likely the
mean drift rate is in the range $100-140\,$arcmin/Gyr, implying that the
average tail star at the end of the observed trailing and leading tail
was lost from the cluster about $3\pm0.5$ and $2.4\pm0.4$ Gyr ago,
respectively.

\subsubsection{Clues for Further Observations of Pal\,5}
%%%%%%%%%%%%%%%%%%%%%%%%%%%%%%%%%%%%%%%%%%%%%%%%%%%%%%%%%%%%%%%%%%%%%%%%%%
In Figure~\ref{fig:proja}, we plot the projection of model A onto the
sky as well as the distribution of stars in their radial-velocity and
distance-modulus offset from the cluster as function of offset in
Galactic longitude. A comparison of the projection of model A with
Fig.~9 of paper~II shows, as already discussed above, that the orbit
chosen in paper~II is presumably not quite correct. This is important
for any application where the orbit geometry is derived from the tidal
tail.

The other two panels of Figure~\ref{fig:proja} plot the observable
radial velocity and the distance modulus for the tail stars versus their
galactic longitude. A measurement of either of these quantities for a
sufficient number of tidal tail stars would constrain the acceleration
in the Galactic halo at Pal\,5's present position and hence directly
give the mass of the Milky Way within $\approx18\,$kpc, a quantity not
very well known.

From this Figure, we see that the distance-modulus offset of the tails
is larger than that of the orbit, simply because of the generic shape of
the tidal tails (see Fig.~\ref{fig:tail:morph}). This enhances the
signal when trying to measure this effect: over the currently observed
arc-length of the tail (from about $-6$ to $4$ in $\Delta(\ell\cos b)$,
see Fig.~9 of paper~II) the distance modulus changes by about
0.25\,mag. The levelling off of the $\Delta m$ at $\ell\sim8^\circ$ is
because this is close to the apogalacticon of the orbit. This implies
that in order to constrain the orbit of Pal\,5, and hence the directly
mass of the Milky Way, distance-modulus determinations should focus on
$\ell\la 8^\circ$.

A similar effect is seen for the radial-velocity offset, which also
shows a larger gradient than the orbit itself. The difference here is
that the spread in $\Delta v_r$ at any fixed position is larger, but
also that gradient is maximal at apogalacticon, in contrast to $\Delta
m$ whose gradient vanishes there. 

Both the gradient in $\Delta m$ and that in $\Delta v_r$ are roughly
proportional to the circular speed $v_c$ of the Galactic halo at the
position of Pal\,5. Thus, in order to determine $v_c$ to 10\% accuracy,
one would need to determine the $\langle v_r\rangle$ at $\sim5^\circ$ on
either side of the cluster to about 4\,km\,s$^{-1}$ accuracy. Note that
more accurate individual measurements are not sensible, because of (1)
the intrinsic width of the tail and (2) possible binarity of the tail
stars. For a determination of $v_c$ to 10\% accuracy from the distance
moduli at, say $\sim5^\circ$ on either side of the cluster, an accuracy
of 0.02\,mag is sufficient. Both these accuracy requirements are not
beyond current capabilities and a determination of the mass of our host
galaxy within 18.5\,kpc to $10-20\%$ accuracy seems possible.

%%%%%%%%%%%%%%%%%%%%%%%%%%%%%%%%%%%%%%%%%%%%%%%%%%%%%%%%%%%%%%%%%%%%%%%%%%
\section{Discussion I:\\ Cluster Evolution Driven by Disk Shocks}
\label{sec:disc:I}
%%%%%%%%%%%%%%%%%%%%%%%%%%%%%%%%%%%%%%%%%%%%%%%%%%%%%%%%%%%%%%%%%%%%%%%%%%
Our simulations were originally intended to understand the evolution and
dynamical status of the globular cluster Palomar 5, but are also
important for understanding disk-shocking driven globular-cluster
evolution in general.

The orbit of Pal\,5 is rather eccentric with an apogalactic radius of
about 19\,kpc, near the cluster's present location. It carries the
cluster to a perigalactic radius as low as 5.5\,kpc, which implies
strong tidal shocks at disk-crossings near perigalacticon. For a
low-concentration low-mass cluster like Pal\,5, these shocks rather than
internal processes (driven by two-body relaxation) dominate the
dynamical evolution \citep{glo99}.

A strong disk shock puts the cluster out of dynamical equilibrium by
unbinding some stars and changing the (cluster-internal) energies of all
the others. In addition to this energy mixing, there is also a general
heating of the cluster. In low-concentration low-mass clusters, like
Pal\,5, this all happens much faster than their dynamical time, i.e.\
the shocks are impulsive. In response to this instantaneous heating, the
cluster expands on its dynamical time scale, then contracts again,
expands and so on. These oscillations are damped, since the internal
dynamical times of the stars differ so that they get out of phase, and
after a few dynamical times the cluster is in a new dynamical
equilibrium.

For very low-concentration clusters the internal dynamical times (1) are
long and (2) do not differ much, so that the damping is less
efficient. This implies that for these systems the settling into a new
equilibrium may be forestalled by the next strong disk shock. Hence,
these clusters will never be in an equilibrium state, but are
continuously rattled by the repeated tidal punches.

\subsection{Evolution of Extended Clusters}
\label{sec:disc:destr}
%%%%%%%%%%%%%%%%%%%%%%%%%%%%%%%%%%%%%%%%%%%%%%%%%%%%%%%%%%%%%%%%%%%%%%%%%%
We find that these shocks can be very efficient in driving the cluster
evolution and may destroy the cluster within a few orbital periods. The
disruption is faster the lower the cluster's concentration and the
larger its size compared to its perigalactic tidal radius. Thus, at
smaller Galactic radii, the lifetime of low-concentration clusters is
lower because of both stronger shocks and shorter orbital periods.

We should emphasize here, that our models were not initially, and never
became, limited by their tidal radii. This implies that globular
clusters extending beyond their theoretical (perigalactic) tidal radius
and orbiting on eccentric orbits on which they experience strong disk
shocks, will not quickly be `tidally stripped' down to their tidal
limit.  This result seems to be at odds with the general wisdom of
globular cluster dynamics in general and with the results of Gnedin et
al.\ in particular. These authors found, using Fokker-Planck
simulations, that in response to disk shocking the clusters become more
compact, which renders the shock-induced evolution self-limiting.

However, we nonetheless think that our results are correct and not
contradicting previous knowledge. Our simulations differ in various ways
from those of Gnedin et al., who consider disk shocks which are
non-impulsive and rare (internal dynamical time much shorter than time
between shocks). We attribute the difference in our findings mainly to
the eccentricity of the orbit of Pal\,5 as compared to that of NGC\,6254
(the cluster studied by Gnedin et al.). After being shock-heated near
perigalacticon, clusters on eccentric orbits approach their new
equilibrium whilst being in the remote parts of their orbit, where the
Galactic tidal field is much smaller (about ten times, see
Fig.~\ref{fig:tfield}) than near perigalacticon, even away from the
disk. As a consequence, the new equilibrium, which does not know
anything about the conditions near perigalacticon, is not limited by the
perigalactic tidal radius.

This mechanism should not be very sensitive to the typical dynamical
time of the cluster, since almost by definition of the tidal radius, the
dynamical time of the outer parts of the cluster never is much smaller
than the orbital time. 

We find that in response to the disk shocks the clusters shrink
somewhat, except for very low-concentration clusters, which even expand,
making them yet more vulnerable and accelerating their destruction.
However, this shrinking is very modest, of the order of at most a few
percent per orbit even for medium-concentration clusters, much too small
to protect them from continued disk shocking. For clusters of higher
concentration and/or on less (or more) eccentric orbits, further studies
are needed to investigate under which conditions tidal-force driven
cluster evolution is self-limiting.

One implication of this possibility of super-tidally limited globular
clusters is that inferring the mass of the Milky Way from the limiting
radii of globular cluster assuming they are tidally limited \citep{K62,
B04} is at best dangerous and may under-estimate the mass of the Milky
Way.

\subsection{Implications for the Globular-Cluster System}
%%%%%%%%%%%%%%%%%%%%%%%%%%%%%%%%%%%%%%%%%%%%%%%%%%%%%%%%%%%%%%%%%%%%%%%%%%
Many studies of the evolution of globular clusters and globular-cluster
systems \citep[e.g.,][]{vh97,bm03,fz01} \emph{assumed\/} that the sizes
of globular clusters are initially upwardly limited by their
(perigalactic) tidal radius. This assumption is based on the idea that
the galactic tidal force field quickly results in such a tidal
limitation, even if the clusters were originally larger. In view of our
results, this assumption appears not generally justified, at least not
for clusters on eccentric orbits. This implies that the above studies
have probably underestimated the importance of disk shocks for the
evolution of globular-cluster systems.

Our simulations predict unambiguously that Pal\,5 will not survive its
next disk crossing in $\sim$\,110\,Myr. If me make the assumption that
there is a plausible initial condition that could have formed 10\,Gyrs
ago and is being disrupted on this orbit now, then the fact that we
observe Pal\,5 within the last per cent of its lifetime, suggests that
we see only the last tip of a melting iceberg. The inner Milky Way may
well have been populated with numerous low-concentration globular
clusters, which by now have all been destroyed, except for those, like
Pal\,5, whose orbits have longer periods and carry them into the more
remote parts of the Galaxy. Currently, we have little possibility of
quantitatively assessing the initial amount of low-concentration and
low-mass globular clusters of the Milky Way galaxy. If tidal tails can
survive for a Hubble time (but see section~\ref{sec:disc:pal5:tail}), we
may be able to identify the debris of destroyed clusters as narrow
streams of stars, which in turn may be identifiable in deep photometric,
astrometric and/or spectroscopic surveys such as SDSS, RAVE and GAIA.

\subsection{Tidal Tails} \label{sec:disc:tail}
%%%%%%%%%%%%%%%%%%%%%%%%%%%%%%%%%%%%%%%%%%%%%%%%%%%%%%%%%%%%%%%%%%%%%%%%%%
Each disk shock triggers the loss of a number of stars: those which
have been accelerated beyond their escape velocity. These stars emerge
with orbits whose energy is either slightly above or below that of the
cluster. This offset in orbital energy is small compared to the orbital
energy of the cluster itself (of the order of one per cent), so that the
stars are not dispersed all over the Galaxy, but stay in a trailing and
leading tidal tail, depending on the sign of the energy offset. This
offset in orbital energy translates into an offset in orbital period and
hence in a drift away from the cluster with a rate of the order of one
per cent, i.e.\ a full wrap-around requires $\mathcal{O}(100)$ periods.

The orbital-energy offsets follow a broad distribution resulting in a
distribution of drift rates. This implies that stars which have been
lost from the cluster at the same time will not stay together but
disperse along the tail. At any distance from the cluster, the tail
contains stars lost at various shock events.

A constant drift rate in conjunction with a constant mass-loss rate
results in a constant linear density of stars in the tidal tail or,
equivalently, to a surface density $\Sigma\propto r^{-\gamma}$ with
$\gamma=1$. This seems to be the generic property of tidal arms that
results from continued tidal shocking \citep{jsh99}. However, when, as a
result of continued mass loss, the cluster mass and hence the escape
velocity drops considerably, stars with ever lower orbital energy
offsets and hence lower drift rate can escape. In this case, the surface
density exponent may be $\gamma>1$, in agreement with our empirical
findings.

%%%%%%%%%%%%%%%%%%%%%%%%%%%%%%%%%%%%%%%%%%%%%%%%%%%%%%%%%%%%%%%%%%%%%%%%%%
\section{Discussion II:\\ The State and Fate of Pal\,5}
\label{sec:disc:II}
%%%%%%%%%%%%%%%%%%%%%%%%%%%%%%%%%%%%%%%%%%%%%%%%%%%%%%%%%%%%%%%%%%%%%%%%%%
\subsection{Is Pal\,5 in Equilibrium?}
%%%%%%%%%%%%%%%%%%%%%%%%%%%%%%%%%%%%%%%%%%%%%%%%%%%%%%%%%%%%%%%%%%%%%%%%%%
Our simulations were primarily aimed at understanding Pal\,5,
predominantly because it is observed to possess a massive tidal
tail. Apart from these tails, the most unusual property of Pal\,5, which
however has not been hitherto noticed, is its enormous size. The cluster
is two times larger than its present theoretical tidal radius and even
four times larger than its perigalactic tidal radius. This property can
in fact be reproduced by (some of) our simulations, and can be
understood in terms of the expansion that immediately follows each
strong disk shock. All our model clusters that can reproduce the present
size (and mass) of Pal\,5 are in a state of near-maximal expansion after
the last disk shock which occurred $\sim$\,146\,Myr ago.

Maximal expansion implies minimal kinetic energy, corresponding to a
small velocity dispersion. Indeed, models that can reproduce cluster
size and mass comparable to Pal\,5 also have the smallest
(line-of-sight) velocity dispersions of all models, namely
$\sigma_{\mathrm{los}}\sim0.45$\,km\,s$^{-1}$. We cannot produce models
with $\sigma_{\mathrm{los}}$ much smaller than that, apparently because
the tidal field limits the lowest velocity dispersions possible for a
surviving cluster on any given orbit.

The line-of-sight velocity distribution of the models is actually not
Gaussian, but contains some high-velocity tails resulting from stars
just leaving the cluster. Correcting for these, we find a Gaussian
$\sigma_{\mathrm{los}}\approx0.38$\,km\,s$^{-1}$ for model A that best
fits size and mass. This value is consistent with our observation for
Pal\,5 of $\sigma_{\mathrm{los}} = 0.22^{+0.19}_{-0.10}\,$km\,s$^{-1}$
(paper~I).

All models that are comparable to Pal\,5 in their cumulative mass
distribution also predict that the cluster will not survive the next
disk crossing in $\sim$\,110\,Myr. The models that best fit the cluster
radial profile have a constant orbit-averaged mass-loss rate of
$\sim5000\,M_\odot$Gyr$^{-1}$. Extrapolating this back in time implies
an original mass of $\sim70000\,M_\odot$, though this may be an
over-estimate, since the strength of the Galactic tidal field was
probably weaker in the past.

\subsection{The Tidal Tail}
%%%%%%%%%%%%%%%%%%%%%%%%%%%%%%%%%%%%%%%%%%%%%%%%%%%%%%%%%%%%%%%%%%%%%%%%%%
Our models also have massive tidal tails. However, models that best fit
the properties of the cluster tend to have a fraction of stars in the
tail within 200\arcmin\ that is about $50-100$\% higher than observed
for Pal\,5. Note however that in this comparison we essentially assumed
implicitly that the cluster's luminosity function equals that of the
tail, which actually appears to be not quite correct \citep{ko04}.

In our best fitting cluster models, $\sim$\,55\% of the tail stars
within 200\arcmin\ are also within 100\arcmin. This corresponds to a
surface density exponent of $\sim$\,1.35 and is consistent with our
observations of Pal\,5 (paper~II).

All our models have a rather smooth density distribution along the tail
with structure of less than $\sim20\%$ in amplitude. This is because, as
mentioned in section \ref{sec:disc:tail} above, even though the stars
are lost at discrete events, they quickly disperse along the tail,
washing out the discreteness of their liberation from the cluster.

This property of the model tails is in clear contrast to the
observations of Pal\,5, in particular its northern tail, which possesses
a substantial clump at about 3\,deg and seems to fade away at
$\sim$\,6\,deg, corresponding to a drift time of only $\sim$\,3\,Gyr
(section~\ref{sec:offset}).

\subsection{What Causes Structure of the Tidal Tail?}
\label{sec:disc:pal5:tail}
%%%%%%%%%%%%%%%%%%%%%%%%%%%%%%%%%%%%%%%%%%%%%%%%%%%%%%%%%%%%%%%%%%%%%%%%%%
Why does the tidal tail of Pal\,5 show this significant structure? 
Obviously, it must be caused by something not accounted for in our
models. Evidently our models lack any Galactic substructure, such as
giant molecular clouds and spiral arms. Clearly, since the Galactic disk
itself is rotating, different parts of the tidal tail encounter
different substructure when crossing the disk. If the tail passes, say,
a giant molecular cloud at a distance smaller than its length, different
parts feel substantially different perturbations and the tail will
develop some substructure.

Another potential source of Galactic substructure that may affect the
tidal tail are clumps in the dark-matter halo, which are actually
predicted by standard $\Lambda$CDM cosmology
\citep[e.g.][]{CDM:1,CDM:2}, or any super massive compact halo objects,
such as proposed by \cite{lo85}. In fact, the very integrity of the tail
of Pal\,5 may impose useful upper limits on the amount, sizes, and
masses of such possible dark substructure in the Milky Way halo.

%%%%%%%%%%%%%%%%%%%%%%%%%%%%%%%%%%%%%%%%%%%%%%%%%%%%%%%%%%%%%%%%%%%%%%%%%%
\section{Summary and Conclusions} \label{sec:summ}
%%%%%%%%%%%%%%%%%%%%%%%%%%%%%%%%%%%%%%%%%%%%%%%%%%%%%%%%%%%%%%%%%%%%%%%%%%
In order to study the disk-shock driven evolution of low-concentration
globular clusters in general and of Pal\,5 in particular, we performed
1056 $N$-body simulations. All these have the same Galactic potential
and orbit (that derived for Pal\,5), but differ in the internal
properties of the cluster: its initial concentration (measured by $W_0$
for our King models), mass, and size (corresponding to our
$R_{\mathrm{t}}$ parameter).

Our simulations demonstrate that disk-shocks near perigalacticon are
very efficient at driving the evolution of low-concentration globular
clusters. In particular, the disk-shocking induced mass loss is
\emph{not\/} self-limiting, but continues at a constant (orbit-averaged)
mass-loss rate without limiting the cluster
\ifpreprint\onecolumn \begin{multicols}{2}\noindent\fi
to its (perigalactic) tidal radius. This behaviour, which is at odds
with previous results by Gnedin et al. (1999), is probably a
consequence of the eccentric orbit, which enables the cluster to
settle into a new equilibrium near apogalacticon, i.e.\ unaffected by
the perigalactic tidal field.

Our findings imply that initially not tidally limited clusters of low to
medium concentration and moving on eccentric orbits may never become
tidally limited and, hence, can be quickly destroyed even when their
relaxation time is long. Clearly, more theoretical and numerical work is
needed to clarify and quantify the issues of cluster destruction by
tides, in particular the effect of the Galactic orbit for the relative
importance of disk shocks in the evolution of globular clusters.

Palomar\,5 is an excellent example for a super-tidal globular cluster
which extends to four times its perigalactic tidal radius. Some of our
simulations actually reproduce this property together with the observed
very low velocity dispersion. Both are the immediate consequences of the
last disk shock, which heated the cluster and made it expand to its
present size. Our simulations also uniquely predict the destruction of
Pal\,5 at its next disk shock in $\sim110\,$Myr. After that the cluster
will loose its identity and eventually only an increasingly separated
pair of tidal tails will remain.

The tidal tails observed within a few degrees from Pal\,5 (paper~II) show
substantial structure which cannot be reproduced by our simulations. We
argued that this structure must be due to Galactic sub-structure (not
accounted for in our simulations), such as giant molecular clouds,
spiral arms, or even dark-matter sub-halos or massive compact halo
objects. More work is needed to quantitatively address the origin of
this structure. 

The tidal tails may also be used to infer the mass of the Milky Way
within 18.5\,kpc to an accuracy of $\sim10\%$ by measuring the radial
velocity or distance modulus of the tail stars over a part of the tidal
tails as large as possible. To this end, an extensive observational
programme is needed to uncover the further extent of the tidal tail
beyond the SDSS data.

\acknowledgements
We thank the anonymous referee for his quick yet useful and detailled
report. EKG gratefully acknowledges partial support from the Swiss
National Science Foundation.

%%%%%%%%%%%%%%%%%%%%%%%%%%%%%%%%%%%%%%%%%%%%%%%%%%%%%%%%%%%%%%%%%%%%%%%%%%
%
% REFERENCES using LaTeX's thebibliography environment
%
%%%%%%%%%%%%%%%%%%%%%%%%%%%%%%%%%%%%%%%%%%%%%%%%%%%%%%%%%%%%%%%%%%%%%%%%%%
\ifpreprint
  \end{multicols}
  \begin{center} \footnotesize
	REFERENCES
  \end{center}
  \vspace*{-7mm}
  \begin{multicols}{2}
  \def\thebibliography#1{%\section*{References}
	\footnotesize
  \list{\null}{\leftmargin1.2em
	\labelwidth0pt
	\labelsep0pt
	\itemindent -1.2em
  	\itemsep0pt plus 0.1pt
	\parsep0pt plus 0.1pt
	\parskip0pt plus 0.1pt
  	\usecounter{enumi}}
  \def\refpar{\relax}
  \def\newblock{\hskip .11em plus .33em
  	minus .07em}
  \sloppy\clubpenalty4000\widowpenalty4000
  \sfcode`\.=1000\relax}
\else  % manuscript
  \clearpage
\fi
%%%%%%%%%%%%%%%%%%%%%%%%%%%%%%%%%%%%%%%%%%%%%%%%%%%%%%%%%%%%%%%%%%%%%%%%%%

%%%%%%%%%%%%%%%%%%%%%%%%%%%%%%%%%%%%%%%%%%%%%%%%%%%%%%%%%%%%%%%%%%%%%%%%%%
%
% TABLE(S), FIGURES AND CAPTIONS (manuscript only)
%
%%%%%%%%%%%%%%%%%%%%%%%%%%%%%%%%%%%%%%%%%%%%%%%%%%%%%%%%%%%%%%%%%%%%%%%%%%

\ifpreprint
	\end{multicols}
\else

\clearpage
	\begin{deluxetable}{llr@{$\,$}r@{$\,$}l}
	%  \tabletypesize{\footnotesize}
	  \tablewidth{0pt}
	  \tablecaption{Time Scales for Pal\,5\label{tab:times}}
	  \tablehead{
	    \colhead{dynamical process} &
	    \multicolumn{4}{c}{time scale}
	    }
	  \tablecolumns{5}
	  \startdata
	%  \tableline \\[-1.7ex]
	  duration of shocks     &                   &$\la$    & 10&Myr\\
	  internal dynamics      &$t_{\mathrm{dyn}}$ &$\sim$   & 80&Myr\\
	  time between shocks    &$t_{\mathrm{disk}}$&$\approx$&300&Myr\\
	  shock-driven evolution &$t_{\mathrm{sh}}$  &$\sim$   &  2&Gyr\\
	  two-body relaxation    &$t_{\mathrm{rh}}$  &$\sim$   & 20&Gyr\\
	  \enddata
	  \tablecomments{Cluster internal time scales are estimated at the
			 half-mass radius $r_{\mathrm{h}}\approx30\,$pc.}
	\end{deluxetable}

\clearpage
	\centerline{\resizebox{90mm}{!}{\includegraphics{Dehnen_fig01}}}
	\figcaption{Orbit of Pal\,5, projected into the meridional
	plane, for the assumed Galactic potential. The current position
	of Pal\,5 is indicated by the star; the trajectory is plotted
	for the last 2.95\,Gyr (\emph{solid}), corresponding to 10
	radial periods, and the future 0.5\,Myr (\emph{dotted}).
	\label{fig:orbit}}

\clearpage
	\centerline{\resizebox{90mm}{!}{\includegraphics{Dehnen_fig02}}}
	\figcaption{Radial profile of the azimuthally averaged surface
	number density of SDSS stars in Pal\,5 and its tails as measured
	in paper~II (\emph{top}), and cumulative number distribution
	averaged over both tails (\emph{bottom}). The typical mass of
	Pal\,5 stars in SDSS is about 0.8\,$M_\odot$.\label{fig:surf}}

\clearpage
	\centerline{\resizebox{90mm}{!}{\includegraphics{Dehnen_fig03}}}
	\figcaption{Strength of the tidal field along the orbit of
	Pal\,5 (Fig.~\ref{fig:orbit}) as measured by the eigenvalues of
	$\partial^2\Phi/\partial x_i\partial x_j$. Positive
	(\emph{solid}) and negative (\emph{dotted}) eigenvalues
	correspond to compressive and stretching tidal forces,
	respectively. The little arrows on the top and bottom of the
	plot indicate moments of disk crossing and pericentric passages,
	respectively. $t=0$ corresponds to today.\label{fig:tfield}}

\clearpage
	\centerline{\resizebox{175mm}{!}{\includegraphics{Dehnen_fig04}}}
	\figcaption{Time evolution of the mass (\emph{left}), velocity
	dispersion (\emph{middle}), and virial radius (\emph{right})
	of the simulated globular cluster as function of time for five
	representative models with parameter $R_{\mathrm{t}}$ and $W_0$
	as indicated and $M_0=12000\,M_\odot$. The \emph{dotted\/}
	vertical lines indicate disk crossings where the the tidal force
	is exceptionally large, quantified by the absolute largest
	eigenvalue of $\partial^2\Phi/\partial x_i\partial x_j$
	exceeding 7000\,(km\,s$^{-1}$\,kpc$^{-1}$)$^2$, see
        Fig.~\ref{fig:tfield}. The
	two simulations with $(R_{\mathrm{t}},W_0)=(11,2.5)$ and
	$(12,2)$ have been stopped before $t=0$, because the number of
        bodies in the cluster dropped below 1000. \label{fig:evolv}}

\clearpage
	\centerline{\resizebox{90mm}{!}{\includegraphics{Dehnen_fig05}}}
	\figcaption{Relative mass-loss and shrinking rates plotted
	versus each other (\emph{top left}) and versus $W_0$ for model
	clusters that have not been dissolved after 2\,Gyr (for models
	dissolved earlier the uncertainties in the shrinking rate can be
	considerable). $\dot{M}$ and $\dot{r}_{\mathrm{vir}}$ have been
	obtained by straight-line fits over the entire time interval,
	excepting the 0.5\,Gyr before cluster dissolution, if
	applicable. For clarity, only models with $M_0=24000\,M_\odot$
	are shown (other values for $M_0$ yield very similar
	results). The symbols refer to different values for
	$R_{\mathrm{t}}$ as indicated.  \label{fig:rates}}

\clearpage
	\centerline{\resizebox{175mm}{!}{\includegraphics{Dehnen_fig06}}}
	\figcaption{Morphology of the simulated tidal tail in the
	simulation with parameters $R_{\mathrm{t}}=10\,$kpc, $W_0=2.75$,
	and $M_0=20000\,M_\odot$ (also called `model A') at various
	times, given in Gyr in each panel. For every tenth body, we plot
	the position (in kpc) projected onto the instantaneous orbital
	plane of the cluster, whose motion is towards the right in each
	panel.  The thin line indicates the cluster
	orbit.\label{fig:tail:morph}}

\clearpage
	\centerline{\resizebox{70mm}{!}{\includegraphics{Dehnen_fig07}}}
	\figcaption{Illustration for the definition of tail coordinates,
	see section~\ref{sec:tail:coords} for details.
	\label{fig:tailcoords}}

\clearpage
	\centerline{\resizebox{140mm}{!}{\includegraphics{Dehnen_fig08}}}
	\figcaption{Distribution of tidal-tail stars over the
	dimensionless drift rate $\eta$, time $t_{\mathrm{l}}$ of loss
	from the cluster, and orbital energy offset $\delta
	E_{\mathrm{{orb}}}$ from the cluster. The values are obtained at
	$t=0$ from model A ($R_{\mathrm{t}}=10\,$kpc, $W_0=2.75$, and
	$M_0=20000\,M_\odot$). Tidal-tail stars with
	$t_{\mathrm{l}}>-0.2\,$Gyr have been omitted. The thin lower
	histogram in the bottom right panel is for stars with
	$t_{\mathrm{l}}\in[-2.5,-2.4]\,$Gyr. In the upper right panel
	only one out of four bodies is plotted. \label{fig:eta}}

\clearpage
	\centerline{\resizebox{90mm}{!}{\includegraphics{Dehnen_fig09}}}
	\figcaption{Plots of $\zeta_\perp$ (\emph{top}) and
	$\zeta_\parallel$ (\emph{bottom}) vs.\ $\eta$ for model A (the
	same as in Fig.~\ref{fig:eta}) at $t=0$ (\emph{left}, near
	apocenter-center) and $t=-0.45\,$Gyr (\emph{right}, near
	pericenter). The thin lines indicate the relations
	(\ref{eq:zeta}) and $\zeta_\perp=0$, which are expected from
	naive modeling. For clarity, only one out of four bodies is
	plotted. \label{fig:zeta}}

\clearpage
	\centerline{\resizebox{90mm}{!}{\includegraphics{Dehnen_fig10}}}
	\figcaption{Distribution of tidal-tail stars in $\tau$ and time
	$t_{\mathrm{l}}$ of loss from the cluster for model A (the
	same as in Fig.~\ref{fig:eta}). The two vertical lines bracket
	the range in $\tau$ inaccessible to stars with
	$t_{\mathrm{l}}<-2.95\,$Gyr. That is, stars lost prior to the
	start of the simulation, which are, of course, not contained in
	our model, contribute only to
	$|\tau|>19\,$Myr. \label{fig:taildens}}

\clearpage
	\centerline{\resizebox{90mm}{!}{\includegraphics{Dehnen_fig11}}}
	\figcaption{For all 695 simulations with a cluster surviving
	until $t=0$, we plot against each other: the average offset in
	orbital energy $\delta E_{\mathrm{{orb}}}$ between tail stars
	and cluster; the average absolute dimensionless drift rate
	$\eta$; and the weighted mean of the initial and final mass of
	the cluster. $\delta E_{\mathrm{{orb}}}$ and $\eta$ have been
	calculated at $t=0$. The thin curves in the two left panels
	follow a power-law with exponent $1/3$. \label{fig:dEorb}}

\clearpage
	\centerline{\resizebox{90mm}{!}{\includegraphics{Dehnen_fig12}}}
	\figcaption{Distribution of the 695 simulated clusters over the
	fractions of stars within projected $16\arcmin$ that are also
	within projected $4\arcmin$ (\emph{x-axis}) and $12\arcmin$
	(\emph{y-axis}). Note that $16\arcmin$ is the limiting radius of
	Pal\,5 itself, whose data are shown as solid square. The open
	square marks the model which best fits the data for Pal\,5 and
	is denoted `model A' in the text. \label{fig:mfrac}}

\clearpage
	\centerline{\resizebox{90mm}{!}{\includegraphics{Dehnen_fig13}}}
	\figcaption{Time evolution of mass (\emph{top}) and velocity
	dispersion (\emph{bottom}) of model A (with
	$R_{\mathrm{t}}=10\,$kpc, $W_0=2.75$, and
	$M_0=20000\,M_\odot$). The dotted vertical lines are
	equivalent to those in Fig.~\ref{fig:evolv}. The cluster is
	destroyed by the first future disk crossing at
	$t=110\,$Myr. \label{fig:simulA}}

\clearpage
	\centerline{\resizebox{90mm}{!}{\includegraphics{Dehnen_fig14}}}
	\figcaption{Distribution of the 695 simulated clusters over the
	final projected mass within 16\arcmin and the line-of-sight
	velocity dispersion within 6\arcmin. Larger symbol sizes
	correspond to models that are closer to Pal\,5 in their enclosed
	projected mass fractions (Fig.~\ref{fig:mfrac}); the largest
	symbol refers to model A. Note that for Pal\,5,
	$M_{r<16\arcmin}=5200\pm700 M_\odot$ and
	$\sigma_{\mathrm{los},r<6\arcmin}=0.22^{+0.19}_{-0.10}\,$
	km\,s$^{-1}$ (assuming a Gaussian distribution, see also Fig.\
	\ref{fig:vrad} and the text for a discussion).
	\label{fig:msig}}

\clearpage
	\centerline{\resizebox{90mm}{!}{\includegraphics{Dehnen_fig15}}}
	\figcaption{\emph{Top}: distribution over distance $r$ from the
	Sun and line-of-sight velocity for stars within projected
	6\arcmin\ of the cluster center for the final state of model A
	(initial parameters: $W_0=2.75$, $R_{\mathrm{t}}=10$, and
	$M_0=20000M_\odot$). \emph{Bottom}: distribution of the same
	stars over $v_{\mathrm{los}}$ and a Gaussian with
	$\sigma_{\mathrm{los}}=0.38\,$km\,s$^{-1}$ (\emph{dotted}).
	\label{fig:vrad}}

\clearpage
	\centerline{\resizebox{90mm}{!}{\includegraphics{Dehnen_fig16}}}
	\figcaption{Distribution of the 695 simulated tails with
	surviving clusters over ($x$-axis) the ratio between mass in the
	tail out to projected $200\arcmin$ and the mass in the cluster
	(within $16\arcmin$) and ($y$-axis) the fraction of tail stars
	within $100\arcmin$ for the trailing (\emph{top}) and leading
	(\emph{bottom}) tail, respectively. The symbol sizes indicate
	how well the model matches the cumulative projected masses of
	the cluster (Fig.~\ref{fig:mfrac}). The solid square represents
	the star-count data for Pal\,5, which have a statistical
	uncertainty of about 0.05 in the $y$ direction. The trailing
	tail of Pal\,5 has a substantial clump at 3 deg from the
	cluster, compromising the measures plotted, in particular that
	on the $y$ axis.  \label{fig:mtail}}

\clearpage
	\centerline{\resizebox{140mm}{!}{\includegraphics{Dehnen_fig17}}}
	\figcaption{Distribution of the 695 simulated tails with
	surviving clusters over ($x$-axis) the mean projected drift rate
	along the orbits and ($y$-axes) the mean (\emph{bottom}) and
	dispersion (\emph{top}) of the rectangular displacement $p$ of
	the tail stars from the orbit. Only tail stars with distance
	(projected arc-length $s$ along the tail) from the cluster within
	$[24\arcmin,\,200\arcmin]$ have been used. The symbol sizes
	indicate how well the model matches the cumulative projected
	masses of the cluster (Fig.~\ref{fig:mfrac}). See text for a
	comparison with Pal\,5 \label{fig:tailoff}}

\clearpage
	\centerline{\resizebox{140mm}{!}{\includegraphics{Dehnen_fig18}}}
	\figcaption{Projection of model A at $t=0$ (today) onto the sky
	in standard Galactic coordinates (\emph{bottom}) and plots of
	the offset in distance modulus (\emph{middle}) and radial
	velocity (\emph{top}) of the stars from the cluster versus
	Galactic longitude. The thin line represents the orbit. Only
	every second body is plotted. \label{fig:proja}}

\fi
%%%%%%%%%%%%%%%%%%%%%%%%%%%%%%%%%%%%%%%%%%%%%%%%%%%%%%%%%%%%%%%%%%%%%%%%%%
\end{document}